\documentclass[11pt,a4paper]{article}
\usepackage{jheppub}
\usepackage{subcaption}

\usepackage{graphicx}
\usepackage{dcolumn}
\usepackage{bm}
\usepackage{amsmath}
\usepackage{mathrsfs}
\usepackage{multirow}
\usepackage{xcolor}
\usepackage{enumitem}

\newcommand{\gsim}{\gtrsim}
\newcommand{\lsim}{\lesssim}

\newcommand{\lf}{\left(}
\newcommand{\ri}{\right)}
\newcommand{\lfb}{\left[}
\newcommand{\rib}{\right]}

\newcommand{\rd}{\partial}
\newcommand{\nn}{\nonumber}

\newcommand{\sqs}{\sqrt{s}}

\newcommand{\rr}{{\gamma\gamma}}
\renewcommand{\lg}{\mathscr{L}}
\newcommand{\m}{\mathscr{M}} 

\newcommand{\mco}{\mathcal{O}}

\newcommand{\br}{\mathcal{B}}
\newcommand{\hc}{{\rm H.c.}}
\newcommand{\sm}{{\rm SM}}
\newcommand{\tot}{{\rm tot}}
\newcommand{\np}{{\rm NP}}

\newcommand{\pb}{{\,{\rm pb}}}

\newcommand{\ifb}{{\,{\rm fb}^{-1}}}

\newcommand{\gev}{{\;{\rm GeV}}}
\newcommand{\tev}{{\;{\rm TeV}}}

\newcommand{\beq}{\begin{equation}}
\newcommand{\eeq}{\end{equation}}
\newcommand{\bea}{\begin{eqnarray}}
\newcommand{\eea}{\end{eqnarray}}
\newcommand{\barr}{\begin{array}}
\newcommand{\earr}{\end{array}}
\newcommand{\bc}{\begin{center}}
\newcommand{\ec}{\end{center}}
\newcommand{\bit}{\begin{itemize}}
\newcommand{\eit}{\end{itemize}}
\newcommand{\ben}{\begin{enumerate}}
\newcommand{\een}{\end{enumerate}}

\newcommand{\al}{\alpha}
\newcommand{\bt}{\beta}

\newcommand{\dt}{\delta}
\newcommand{\Dt}{\Delta}

\newcommand{\sg}{\sigma}

\newcommand{\kp}{\kappa}

\newcommand{\gm}{\gamma}
\newcommand{\Gm}{\Gamma}
\newcommand{\lm}{\lambda}

\newcommand{\mch}{M_{H^\pm}}

\newcommand{\tb}{t_\beta}
\newcommand{\cb}{c_\beta}
\renewcommand{\sb}{s_\beta}

\newcommand{\cba}{c_{\beta-\alpha}}
\newcommand{\sba}{s_{\beta-\alpha}}

\newcommand{\ma}{M_A}

\newcommand{\mh}{m_{h}}
\newcommand{\mhh}{M_{H}}
\newcommand{\mtp}{M_{t'}}
\newcommand{\mbp}{M_{b'}}
\newcommand{\mnup}{M_{\nu'}}
\newcommand{\mtaup}{M_{\tau'}}

\newcommand{\ee}      {{e^+ e^-}}
\newcommand{\mmu}      {{\mu^+ \mu^-}}
\newcommand{\ttau}     {{\tau\tau}} 
\newcommand{\ttaup}      {{\tau'\tau'}}

\newcommand{\ttop}      {{t\bar{t}}}
\newcommand{\ttopp}      {{t'\bar{t}'}}
\newcommand{\bb}      {{b \bar{b}}}
\newcommand{\bbp}      {{b' \bar{b}'}}
\newcommand{\qq}      {{q \bar{q}}}

\renewcommand\epsilon{\varepsilon}

\newcommand{\vtw}[2]{\left(\begin{array}{c}#1 \\ #2 \end{array}\right)}

\preprint{
\begin{flushright}
CTPU-PTC-18-31
\end{flushright}
}

\title{
Confronting the fourth generation two Higgs doublet model
with the phenomenology of heavy Higgs bosons}

\author[a]{Sin Kyu Kang,}
\author[b]{Zhuoni Qian,}
\author[c]{Jeonghyeon Song,}
\author[c]{and Yeo Woong Yoon}
\affiliation[a]{School of Liberal Arts, Seoul-Tech. Seoul 01811,  Korea}
\affiliation[b]{Center for Theoretical Physics of the Universe, Institute for Basic Science (IBS), Daejeon, 34126, Korea}
\affiliation[c]{Department of Physics, Konkuk University, Seoul 05029, Korea}

\emailAdd{skkang@snut.ac.kr}
\emailAdd{zhuoniq@ibs.re.kr}
\emailAdd{jhsong@konkuk.ac.kr}
\emailAdd{ywyoon@kias.re.kr}

\abstract{
A sequential fourth generation is known to be excluded 
because the non-decoupling contribution to $\kappa_g$,
the Higgs coupling modifier with a gluon pair,
is unacceptably large.
Recently a new way to save the model
was suggested in the Type-II two Higgs doublet model:  
if the Yukawa couplings of down-type fermions have wrong-sign,
the contributions from $t'$ and $b'$ to $\kappa_g$ are cancelled.
We study the theoretical and experimental constraints on this model, focusing on the heavy Higgs bosons. 
Two constraining features are pointed out.
First the exact wrong-sign limit does not allow the alignment,
which makes 
the perturbative unitarity for the scalar-scalar scattering 
put the upper bounds on the heavy Higgs boson masses like
$M_H, M_A \lesssim 920$ GeV and $M_{H^\pm} \lesssim 620$ GeV.
Secondly, the Yukawa couplings of the fourth generation fermions to the heavy Higgs bosons
are generically large as being proportional to the heavy fermion mass 
and, for the down-type fermions, to $\tan\beta$ as well.
The gluon fusion productions of $H$ and $A$ through the fourth generation quark loops
become significant.
We found that the current LHC data on 
$pp \to Z Z$ for $H$
along with the theoretical and indirect constraints
exclude the model at leading order.
}

\begin{document}

\maketitle

\section{Introduction}
Our dearest wish for new physics (NP) beyond the standard model (SM) 
has not been fulfilled yet 
since the dedicated searches for new particles at the LHC
found no new signals and various precision data 
are consistent with the SM predictions.
The usual strategy for a NP theory is to hide it in the decoupling limit:
very heavy new particles cannot be observed at the current 13 TeV LHC
nor significantly contribute to the precision data including the Higgs signals.
One important exception is a sequential fourth generation model
where the new chiral fermions ($t'$, $b'$, $\nu'$ and $\tau'$) acquire their masses via the same Higgs mechanism.
This model addresses one of the most fundamental questions$-$whether there exist
only three fermion generations in the universe~\cite{Frampton:1999xi,Choudhury:2001hs,Bobrowski:2009ng}.
The discovery of the SM-like Higgs boson with a mass of 125 GeV~\cite{Aad:2012tfa,Chatrchyan:2012xdj}
excludes this model
since the contributions of the heavy fourth generation quarks to the gluon fusion production of the Higgs boson
do not decouple but saturate to a constant value which is unacceptably large.

Based on the observation that the NP contribution to 
the Higgs coupling modifier with a gluon pair, $\kp_g$, is proportional to the sum of the Higgs coupling modifiers 
of $t'$ and $b'$, i.e.,  $(\kp_{t'}+ \kp_{b'})$ 
when $\mtp,\mbp \gg \mh$,
the authors in Ref.~\cite{Das:2017mnu}
found that the sequential fourth generation model 
can survive the Higgs precision constraint
if the down-type quark Yukawa coupling has opposite sign to that of the up-type quark,
i.e., $\kp_{b'}=-\kp_{t'}$.
More interesting is that
in the exact wrong-sign limit where all of the down-type fermions have opposite Higgs coupling 
to the up-type fermion,
the new contributions to
$\kp_\rr$ and $\kp_{Z\gm}$
also vanish.
The wrong-sign Yukawa couplings for the down-type fermions
cannot be realized in the SM with only one Higgs doublet~\cite{Bhattacharyya:2012tj}:
the Higgs sector should be extended as 
in the two Higgs doublet model (2HDM)~\cite{Carmi:2012yp,Chiang:2013ixa,Ferreira:2014naa,Fontes:2014tga}.
Authors in Ref.~\cite{Das:2017mnu} showed that the Type-II 2HDM with a sequential fourth generation
satisfies not only the Higgs signal strength measurements 
but also the oblique parameters $\Dt S$ and $\Dt T$.
Follow-up studies focused on top quark dipole moments~\cite{Hernandez-Juarez:2018uow},
dark matter~\cite{Han:2017etg}, and lepton flavor changing in the Higgs boson decays~\cite{Chamorro-Solano:2017toq}.

Albeit the appeal of the model in satisfying the Higgs precision constraint 
through such a simple remedy,
this model has a potentially dangerous spot, 
the phenomenology of the heavy Higgs bosons.
The 2HDM has five physical Higgs bosons,
the light CP-even scalar $h$,
the heavy CP-even scalar $H$, the CP-odd pseudoscalar $A$,
and two charged Higgs bosons $H^\pm$.
With a \emph{sequential} fourth generation,
the Yukawa couplings of the fourth generation fermion $F$ with all the Higgs bosons
are proportional to the fermion mass $M_F$, which are naturally very large.
In addition, the down-type fermion Yukawa couplings with $H$ and $A$, $Y^{H/A}_{b'}$,
are proportional to $\tan\beta$ while the up-type fermion couplings are inversely proportional to $\tan\beta$,
where $\tan\beta$ is the ratio of two vaccum expectation values (VEVs) of two Higgs doublet fields.
The cancellation of $t'$ and $b'$ contributions to the $hgg$ coupling does not occur 
to the $Hgg$ and $Agg$ couplings.
At the LHC, $H$ and $A$ can be copiously produced.

Another concern is from the exact wrong-sign limit, 
the key that allows the sequential fourth generation.
It relates two mixing angles $\alpha$ and $\beta$ through
$\alpha+ \beta ={\pi}/{2}$, 
where $\alpha$ is the mixing angle between $h$ and $H$.
Unless $\tan\beta$ becomes very large, the exact wrong-sign limit cannot approach the alignment limit of $\sin(\beta-\alpha)=1$.
Since too large $\tan\beta$ breaks the perturbativity of $Y^{H/A}_{b'}$,
the exact wrong-sign limit brings about significant deviation from the alignment limit.
One important consequence is the different dependence of the scalar quartic couplings on the heavy Higgs bosons 
masses from the alignment limit.
As shall be shown,
this difference causes a serious result on the 
perturbative unitarity of scalar-scalar scattering,
the \emph{upper} bounds on the heavy Higgs boson masses.
The decoupling limit cannot be achieved along with the exact wrong-sign limit.
The model does not have a safety zone.
We need a comprehensive study on the phenomenology of the model including the heavy Higgs bosons,
which is our main purpose.

This paper is organized as follows.
In Sec.~\ref{sec:review},
we review the Type-II 2HDM with a sequential fourth generation in the exact wrong-sign limit.
The Higgs coupling modifiers in this model are compared with those in the alignment limit,
with special focus on $\kp_V$.
Section \ref{sec:theoretical} deals with the theoretical constraints on the scalar sector:
the bounded-from-below potential, unitarity, perturbativity, and vacuum stability.
Here we shall explicitly show that the behavior of $\lm_3$ about $\mhh$
is very different from that in the alignment limit. 
In Sec.~\ref{sec:indirect},
we study the indirect constraints from the electroweak oblique parameters 
and the Higgs precision data.
The observed $\kp_V$ restricts $\tan\beta$ significantly.
Based on the narrowed parameter space from the theoretical and indirect constraints,
we study the decay and production of $H$ and $A$ in Sec.~\ref{sec:H:A}.
Here we shall point out that above the $\bbp$ threshold, 
the total decay widths of both $H$ and $A$ become so wide like $\Gm_\tot^{H/A} \sim M_{H/A}$.
The ordinary analysis based on $\sg \times \br$ does not work here.
We suggest a method to probe this very fat resonance, not relying on the total event counting.
The production cross sections of the gluon fusion of $H$ and $A$ are also studied.
In Sec.~\ref{sec:direct:search},
we consider direct search results for new particles at the LEP and LHC, 
$\ee \to 4b, 4\tau, 2b2\tau$ and $pp \to \ttau, Z Z, Zh$.
Two smoking-gun signals, $Zh$ for $A$ and $Z Z$ for $H$,
are to be elaborated, drawing a rather strong conclusion
that this model at leading order is excluded by the combination of theoretical and experimental constraints.
In Sec.~\ref{sec:conclusions}, we summarize and conclude.

\section{Review of the 2HDM-SM4}
\label{sec:review}

We consider a 2HDM with a sequential fourth generation
in the exact wrong-sign limit.
The Higgs sector and the fermion sector are extended by introducing
two complex $SU(2)_L$ Higgs doublet scalar fields, $\Phi_1$ and $\Phi_2$~\cite{Branco:2011iw},
and a 
sequential fourth generation respectively:
\bea
\label{eq:fields}
\Phi_i = \left( \begin{array}{c} w_i^+ \\
\dfrac{v_i +  h_i + i \eta_i }{ \sqrt{2}} 
\end{array} \right),
\quad
\vtw{t'_L}{b'_L}, \quad t'_R, \quad b'_R, \quad  \vtw{\nu'_L}{\tau'_L},\quad \nu'_R, \quad \tau'_R, 
\eea
where $i=1,2$, and $v_{1,2}$ is the nonzero VEV of $\Phi_{1,2}$.
Note that the anomaly cancellation condition~\cite{Adler:1969gk,Bell:1969ts} 
requires the existence of the fourth generation leptons.
When parametrizing $\tb =v_2/v_1$,
one linear combination $H_1 = \cb \Phi_1 + \sb \Phi_2$
has nonzero VEV of $v =\sqrt{v_1^2+v_2^2}=246\gev $,
which generates the electroweak symmetry breaking.
Its orthogonal combination $H_2 = -\sb \Phi_1 +\cb \Phi_2$
acquires zero VEV.
For simplicity of notation, we take $s_x=\sin x$, $c_x = \cos x$, and $t_x = \tan x$.

In order to avoid flavor changing neutral current (FCNC) at tree level,
a discrete $Z_2$ symmetry is imposed, under which $\Phi_1 \to \Phi_1$
and $\Phi_2 \to -\Phi_2$~\cite{Glashow:1976nt,Paschos:1976ay}.
Then the most general scalar potential with CP invariance and softly broken $Z_2$ symmetry is  
\bea
\label{eq:V}
V_\Phi = && m^2 _{11} \Phi^\dagger _1 \Phi_1 + m^2 _{22} \Phi^\dagger _2 \Phi_2 
-m^2 _{12} ( \Phi^\dagger _1 \Phi_2 + \hc) \nn \\
&& + \frac{1}{2}\lambda_1 (\Phi^\dagger _1 \Phi_1)^2 
+ \frac{1}{2}\lambda_2 (\Phi^\dagger _2 \Phi_2 )^2 
+ \lambda_3 (\Phi^\dagger _1 \Phi_1) (\Phi^\dagger _2 \Phi_2) 
+ \lambda_4 (\Phi^\dagger_1 \Phi_2 ) (\Phi^\dagger _2 \Phi_1) \nn\\
&& + \frac{1}{2} \lambda_5 
\left[
(\Phi^\dagger _1 \Phi_2 )^2 +  \hc
\right] . \label{eq:potential}
\eea
The CP invariance requires all of the parameters to be real.
Note that the soft $Z_2$ symmetry breaking parameter $m_{12}^2$ can be negative.

There are five physical Higgs bosons,
the light CP-even scalar $h$,
the heavy CP-even scalar $H$, the CP-odd pseudoscalar $A$,
and two charged Higgs bosons $H^\pm$.
They are related with the weak eigenstates in Eq.~(\ref{eq:fields}) via
\bea
\left(
\begin{array}{c}
h_1 \\ h_2
\end{array}
\right) =
\mathbb{R}(\al) 
\left(
\begin{array}{c}
H \\ h
\end{array}
\right),
\quad
\left(
\begin{array}{c}
\eta_1 \\ \eta_2
\end{array}
\right) =
\mathbb{R}(\beta) 
\left(
\begin{array}{c}
z^0 \\ A
\end{array}
\right)
, \quad
\left(
\begin{array}{c}
w_1^\pm \\ w_2^\pm
\end{array}
\right) =
\mathbb{R}(\al) 
\left(
\begin{array}{c}
w^\pm \\ H^\pm
\end{array}
\right),
\eea
where $z^0$ and $w^\pm$ are the Goldstone bosons to be eaten by the $Z$ and $W$ bosons, respectively.
The rotation matrix $\mathbb{R}(\theta)$ is
\bea
\mathbb{R}(\theta) = \left(
\begin{array}{cr}
c_\theta & -s_\theta \\ s_\theta & c_\theta
\end{array}
\right).
\eea
We consider the normal scenario where the observed Higgs boson is the lighter CP-even $h$,
although the other scenario with $\mhh=125\gev$ is still
allowed by the Higgs precision data~\cite{Chang:2015goa,Bernon:2015wef}.

The Yukawa couplings are different according to the $Z_2$ parity of the fermions.
We fix that $Q_L \to Q_L$ and $L_L \to L_L$ under the $Z_2$ symmetry,
where $Q_L$ and $L_L$ are the left-handed quark and lepton doublets, respectively.
Then each right-handed fermion field only couples to one scalar doublet field.
There are four different ways to assign the $Z_2$ symmetry on the right-handed fermion fields,
leading to four different types in the 2HDM, Type-I, Type-II, Type-X, and Type-Y.
We parameterize the Yukawa interactions with the neutral Higgs bosons as
\bea
\label{eq:Lg:Yuk}
-\lg_{\rm Yuk} 
&=&\sum_{f} \frac{m_f}{v} 
\lf \kp_f \bar{f}f h +  \xi^H_f \bar{f}f H  -i \xi^A_f \bar{f} \gm_5 f A \ri .
\eea
Note that $\kp_f$ is the Higgs coupling modifier,
parameterizing the NP effects on the Higgs couplings:
\bea
\label{eq:kappa:def}
\kp_i  =\frac{g_{iih}}{g_{iih}^\sm}  \equiv 1+ \dt \kp_i.
\eea
While $\kp_f$'s are different according to the 2HDM type, 
$\kp_V$ and $\xi_V$ ($V=W^\pm,Z$)
have the common leading order expressions of
\bea
\label{eq:kappaV:tree}
\kp_V = \sba, \quad \xi_V = \cba.
\eea

Because the observed Higgs boson at a mass of 125 GeV is very SM-like,
the so-called alignment limit is usually adopted in the 2HDM\footnote{The terminology \emph{alignment} was 
first used when tuning the Yukawa sector to avoid the tree-level FCNC 
without introducing the $Z_2$ symmetry~\cite{Pich:2009sp}. 
}, defined by
\bea
\label{eq:alignment}
&& \al =\bt -  \frac{\pi}{2} \quad\hbox{ (alignment limit) }
\\ \nn
\longrightarrow &&  \kp_u = \kp_d =1, \quad \kp_V = 1,
\quad
\xi_V =0,
\eea
where $u=t,t',\nu'$ and $d=b,b',\tau,\tau'$.
With a sequential fourth generation, however,
this alignment limit does not guarantee a SM-like Higgs boson 
because of the large contribution from the fourth generation fermions
to the loop induced couplings of the Higgs boson,
especially to the $\kp_g$:
\bea
\label{eq:kpg}
\kp_g 
= \frac{\kp_t A^h_{1/2}(\tau_t) + \sum_{F} \kp_F A^h_{1/2}(\tau_F)  }
{A^h_{1/2}(\tau_t)},
\eea
where $\tau_f = \mh^2/4 m_f^2$, 
$F=t',b'$, and 
the expression for the loop function $A^h_{1/2}(\tau)$ is referred to Ref.~\cite{Djouadi:2005gi}.
It is known that $A^h_{1/2}(\tau_f)$ approaches the value of $4/3$ when $m_f \gg \mh$.
In the alignment limit ($\kp_{t'}=\kp_{b'}=1$) with $M_F \gg \mh$,
therefore,
the value of $\kp_g$ approaches 3.
We cannot but conclude that a sequential fourth generation 
in the SM or the aligned 2HDM is excluded by the Higgs precision data.

Based on the observation that $\dt \kp_g$ is proportional to $(\kp_{t'}+ \kp_{b'})$ 
for $M_F \gg \mh$
and the current LHC data cannot determine the sign of $\kp_b$ yet,
the exact wrong-sign limit~\cite{Das:2017mnu,Fontes:2014tga,Ferreira:2014dya}
is suggested, given by
\bea
\label{eq:exact wrong-sign:def}
&& \al = \frac{\pi}{2} - \bt \quad \hbox{ (exact wrong-sign limit) }
\\ \nn
\longrightarrow &&  \kp_u = 1, \quad \kp_d =-1, \quad \kp_V =\frac{\tb^2-1}{\tb^2+1},
\quad
\xi_V = \frac{2 \tb}{\tb^2+1}.
\eea
The wrong-sign Yukawa couplings for  the down-type fermions
cannot be realized in the SM where there exists only one scalar doublet field:
all of the Yukawa couplings can be set positive by chiral rotation.
We need an additional Higgs doublet field,
which can be minimally realized in the 2HDM.
Among four types of the 2HDM, only Type-II can accommodate the exact wrong-sign limit
where both $b'$ and $\tau'$ have opposite Yukawa couplings to $t'$ and $\nu'$.
In what follows,
2HDM-SM4 denotes the Type-II 2HDM with a sequential fourth generation in the exact wrong-sign limit.

More surprising feature of the exact wrong-sign limit
is that new contributions from the sequential fourth generation fermions to $\kp_\rr$ and $\kp_{Z\gm}$
are also suppressed in the heavy $M_F$ limit~\cite{Das:2017mnu}:
\bea
\label{eq:NP:kappa:g:gm}
\dt \kp_\rr &\propto& \sum_{f=t',b',\tau'} Q_f^2 N_C^f \kp_f = 0,
\\ \nn
\dt \kp_{Z\gm} &\propto& \sum_{f=t',b',\tau'} Q_f (T_3^f)_L N_C^f \kp_f = 0,
\eea
where 
$Q_f$ is the electric charge of the fermion $f$,
$N_C^f$ is the color factor, 
and  $(T_3^f)_L$ is the isospin projection of the left-handed $f_L$.

In the 2HDM, however, there exist other Higgs bosons, $H$, $A$, and $H^\pm$.
The exact wrong-sign condition simplifies $\xi^{H,A}_{u,d}$, defined in Eq.~(\ref{eq:Lg:Yuk}), into
\bea
\label{eq:exact wrong-sign:xi}
\xi^H_u = \xi^A_u = \frac{1}{\tb} \equiv \xi_u,
\quad
\xi^H_d = \xi^A_d = {\tb} \equiv \xi_d.
\eea
An immediate concern is that the Yukawa couplings of $b'$ and $\tau'$ with $H$ and $A$,
proportional to the heavy fermion masses,
can be dangerously large especially in the large $\tb$ limit.
Theoretical principles and collider experiments associated with $H$ and $A$
shall constrain the model significantly.

We now specify the model parameters in the 2HDM-SM4.
In the scalar potential sector,
there are 7 free parameters of 
$m_{12}^2$, $\tb$, and $\lm_{1,\cdots,5}$,
after applying the tadpole conditions for $m_{11}$ and $m_{22}$.
Equivalently we can take the physical parameters of 
$\mh$, $\mhh$, $\ma$, $\mch$,  $m^2$,
$\al$ and $\beta$,
where $m^2 = m_{12}^2/(\sb\cb)$ is chosen because of its efficiency to show the invariance
under the reparameterization in the space of Lagrangian~\cite{Ginzburg:2004vp}. 
Since
$\mh=125\gev$ is known
and the exact wrong-sign limit relates $\al$ and $\beta$ as $\al+\bt=\pi/2$,
there are five free parameters in the scalar sector.
In the fourth generation fermion sector, only their masses are unknown 
because their gauge and Yukawa couplings are the same as the SM fermions.
In summary, the 2HDM-SM4 has the following model parameters:
\bea
\label{eq:model:parameters}
\tb,  \quad m^2, \quad \mhh,\quad \ma, \quad \mch,\quad \mtp, \quad \mbp, \quad \mtaup,
\quad \mnup.
\eea

Brief comments on the fourth generation fermion mass $M_F$ and $\mch$ are in order here. 
For $M_F$, 
there are two kinds of constraints working in the opposite way,
one from the unitarity and the other from the direct searches.
First, 
the perturbative unitarity for the fermion-fermion scattering puts the upper mass bound
as $m_{q'} \lsim 550\gev$ and $m_{\ell'}\lsim 1.2\tev$~\cite{Dighe:2012dz,Dawson:2010jx}.
On the other hand, the direct searches for $t'$ and $b'$ at the LHC 
put the lower bounds of
$\mtp \gsim 680\gev$
under the assumption that the produced $t'$ and $b'$ decay into a SM quark accompanied 
by a $W$ or $Z$ boson~\cite{Chatrchyan:2012fp}.
If the mixing between the SM quarks and the fourth generation quarks is extremely small like $V_{i4} \lsim 10^{-7}$, however,
no limits can be set~\cite{Chatrchyan:2012fp}.
The CDF collaboration took the assumption of specific flavor-mixing rates
and put the lower bound on $\mtp,\mbp \gsim 335-385 \gev$~\cite{Lister:2008is,Aaltonen:2009nr}.
With general flavor mixing, 
the lower mass bounds were recalculated to be as low as 290 GeV~\cite{Flacco:2010rg,Flacco:2011ym}.
The fourth generation leptons have weaker bounds as
$m_{\tau'} > 100.8\gev$ and $m_{\nu'} > 41\gev$~\cite{Patrignani:2016xqp}.
The charged Higgs boson mass in a Type-II 2HDM is most strongly constrained 
by the FCNC process  $\bar{B} \to X_s \gm$.
The updated next-to-next-to-leading order SM prediction 
of $\br_\sm(\bar{B} \to X_s \gm)$~\cite{Misiak:2017bgg,Misiak:2017zan}
and the recent Bell result~\cite{Belle:2016ufb}
got closer, yielding
$\mch > 570 ~(440) \gev$ for $\tb\gsim 2$ at 95\% (99\%) C.L.
For $\tb\lsim 2$, the lower bound on $\mch$ increases significantly.
Note that the fourth generation quarks do not affect the process $\bar{B} \to X_s \gm$
under the assumption of $V_{4i} \lsim 10^{-7}$.

\section{Theoretical constraints on the scalar potential}
\label{sec:theoretical}

The quartic coupling constants in the scalar potential $V_\Phi$
can be rewritten in terms of the physical mass parameters,
which are in the exact wrong-sign limit
\bea
\lm_1 &=& \frac{1}{v^2 }
\left[
\tb^2 \lf \mhh^2 -m^2 \ri + m_h^2 
\right],
\\ \nn
\lm_2 &=&
\frac{1}{v^2 }
\left[
 \frac{1}{\tb^2} \lf \mhh^2 -m^2 \ri  + m_h^2 
\right],
\\ \nn
\lm_3 &=& \frac{1}{v^2}
\left[
-m^2  +\mhh^2-m_h^2 + 2 \mch^2
\right],
\\ \nn
\lm_4 &=& \frac{1}{v^2}
\left[
m^2 + \ma^2 - 2 \mch^2
\right],
\\ \nn
\lm_5 &=& \frac{1}{v^2}
\left[
m^2 - \ma^2 
\right].
\eea
They are constrained by the following theoretical conditions:
\begin{enumerate}
\item The scalar potential $V_\Phi$ should be bounded from below in any direction,
requiring~\cite{Ferreira:2014sld,Ivanov:2006yq}
\bea
\lambda_1 > 0, \quad 
\lambda_2 > 0, \quad \lambda_3 > -\sqrt{\lambda_1 \lambda_2}, \quad  
\lambda_3 + \lambda_4 - |\lambda_5| > - \sqrt{\lambda_1 \lambda_2}.  \label{eq:stability} 
\eea
\item The tree-level perturbative unitarity demands~\cite{Lee:1977eg,Kanemura:2015ska,Arhrib:2000is}
\bea
\left| 
a_{i,\pm}
\right| \leq 1,
\eea
where $a_{i,\pm}$ ($i=1,\cdots,6$) are 
the eigenvalues of the $ T$ matrix for the $S$-wave amplitudes 
of the scalar-scalar scattering, given by
\bea
\label{eq:unitarity}
a_{1,\pm} & = & \frac{1}{32\pi} 
\lfb
3(\lambda_1+\lambda_2)\pm  \sqrt{9(\lambda_1-\lambda_2)^2+4(2\lambda_3+\lambda_4)^2}
\rib,  \\ \nn
a_{2,\pm} & = &
\frac{1}{32\pi} 
\lfb
 \lambda_1+\lambda_2 \pm \sqrt{(\lambda_1-\lambda_2)^2+4\lambda_4^2} 
\rib
,  \\ \nn
a_{3,\pm} & = & 
\frac{1}{32\pi} 
\lfb
 \lambda_1+\lambda_2 \pm \sqrt{(\lambda_1-\lambda_2)^2+4\lambda_5^2} \rib, \nn \\
a_{4,\pm} &=&  
\frac{1}{16\pi} 
\lf
 \lambda_3 +2\lambda_4 \pm 3\lambda_5\ri,
\\ \nn
a_{5,\pm} &=& 
\frac{1}{16\pi} 
\lf
\lambda_3\pm \lm_4
\ri,
\\  \nn
a_{6,\pm} &=&
\frac{1}{16\pi} 
\lf
\lambda_3\pm \lm_5
\ri.
\eea
\item The perturbativity of scalar quartic couplings requires
\bea
\label{eq:perturbativity}
\left| \lm_{i} \right| < 4 \pi, \quad i=1,\cdots,5.
\eea
\item The vacuum of $V_\Phi$ should be global, which happens if and only if~\cite{Barroso:2013awa}
\bea
D = m_{12}^2 \left( m_{11}^2 - k^2 m_{22}^2 \right) (\tb - k) >0,
\eea
where $k=(\lm_1/\lm_2)^{1/4}$.
\end{enumerate}

In the exact wrong-sign limit, the theoretical constraints are more difficult to satisfy than in the alignment limit.
Crucial is $\lm_3$,
which shows different dependence on $\mhh$:
\bea
\label{eq:lm3}
\lm_3  = 
\left\{
	\begin{array}{ll}
	\dfrac{1}{v^2} \lfb 2 \mch^2 +\mhh^2 -m_h^2  -m^2  \rib , & \hbox{ in the exact wrong-sign limit;} 
	\\[15pt]
	\dfrac{1}{v^2} \lfb 2 \mch^2 -\mhh^2 +m_h^2  -m^2 \rib, & \hbox{ in the alignment limit.}
	\end{array}
\right.
\eea 
In both limits, $\lm_3$ has the same dependence on $\mch^2$
but has opposite signs for the $\mhh^2$ terms. 
Since $\mch$ in a Type-II 2HDM should be very heavy to explain the $b\to s \gm$ result,
heavy $\mhh$ in the exact wrong-sign limit shall easily increase $\lm_3$ above $4\pi$.
In the alignment limit,
the negative $\mhh^2$ contribution  cancels the positive $\mch^2$ contribution to some extent,
being able to control the $\lm_3$ value.

\begin{figure}[t!]
\begin{center}
\includegraphics[height=7cm]{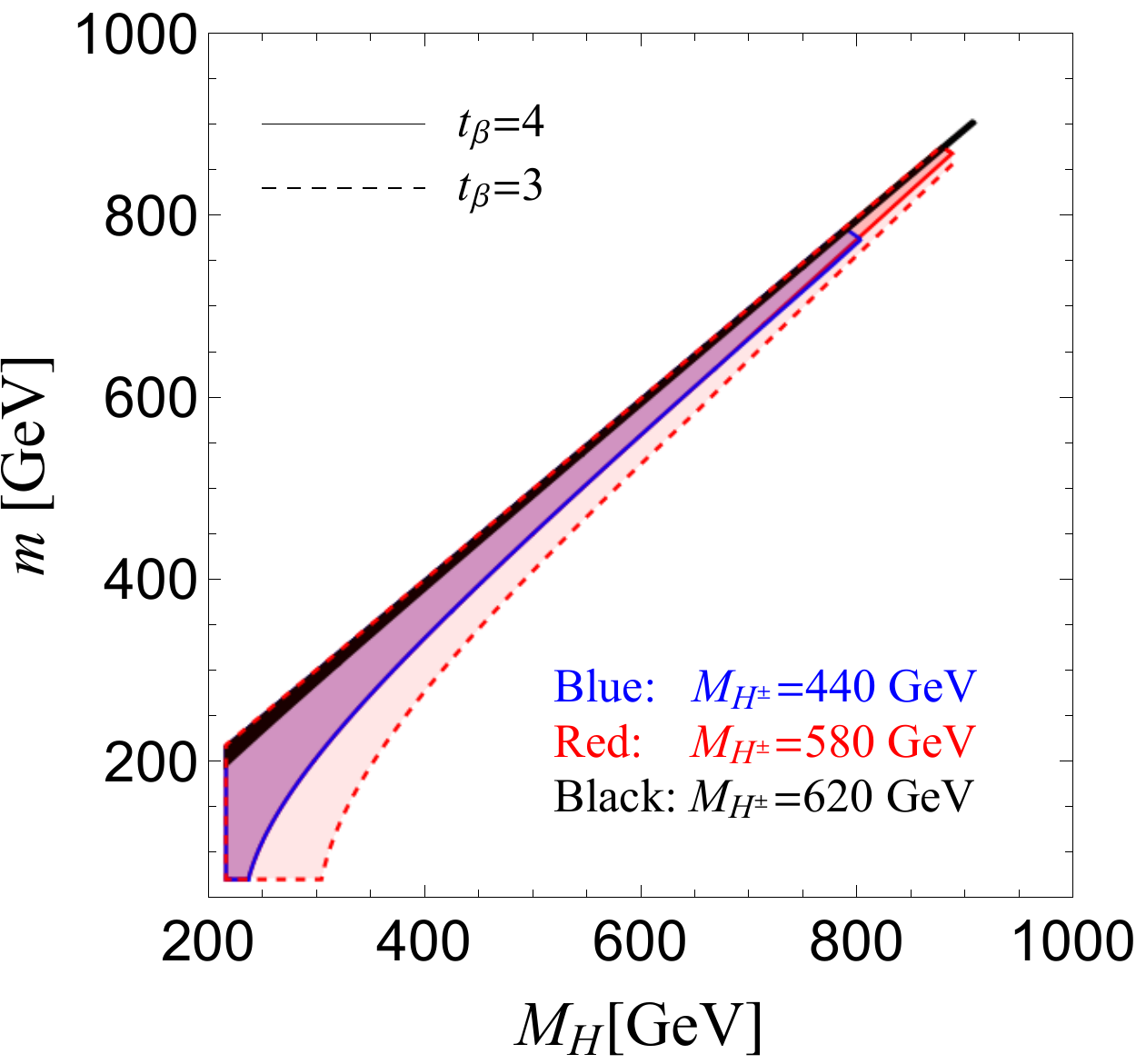}
\includegraphics[height=7cm]{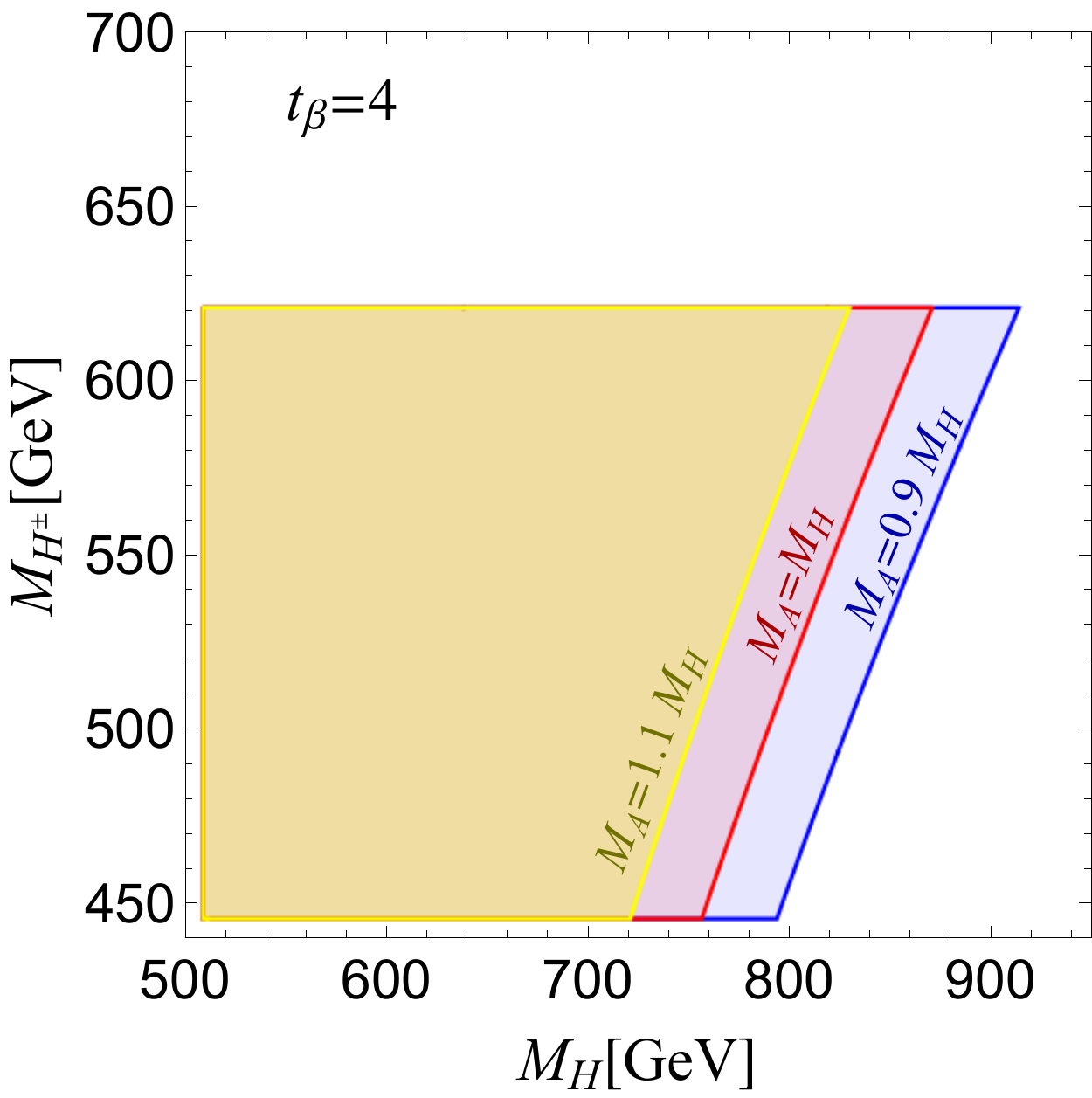}
\caption{\label{fig-theoretical}
\baselineskip 3.0ex
Left panel: Theoretically allowed parameter space of $(M_H,m)$ for $\tb=3,4$, $\mch=440,~580,~620\gev$, 
and $\ma=0.9\,\mhh$. 
Right panel: 
Theoretically allowed parameter space of  $(\mhh,\mch)$ for $\tb=4$ and
$\ma=(1\pm 0.1)\mhh$.
}
\end{center}
\end{figure}

Figure \ref{fig-theoretical}
presents the theoretically allowed parameter space.
In the left panel, 
we show the allowed parameter space\footnote{Since only positive $m^2$ is allowed by the theoretical constraints,
$m=\sqrt{m^2}$ is presented.} of  $(\mhh,m)$
for $\mch=440\gev$ and $\tb=4$ (blue region), $\mch=580\gev$ and $\tb=3,4$ (red region),
and $\mch=620\gev$ and $\tb=4$ (black region).
We also set $\ma=0.9 \mhh$ 
in order to explain the oblique $\Dt T$ parameter.
Two cases of $\tb=3$ and $\tb=4$ for the same $\mch=580\gev$ show that
small $\tb$ allows more freedom for $m$, but does not significantly change the allowed range for $\mhh$.
When $\tb$ is very large,
only a fine linear line along $\mhh=m$ is allowed, irrespective of $\mch$.
The most important result is that 
theoretical constraints put an \emph{upper} bound on $\mhh$,
of which the $\mch$ dependence is small and the $\tb$ dependence is negligible. 

The charged Higgs boson mass is also bounded from above.
In the right panel of Fig.~\ref{fig-theoretical},
we show the theoretically allowed parameter space of $(\mhh,\mch)$
for $\tb=4$.
We consider three cases of $\ma=\mhh$,  $\ma=1.1\,\mhh$, and $\ma=0.9\,\mhh$
to explain the oblique parameter $\Dt T$.
It can be seen that $\mhh$ cannot exceed about 920 GeV
and $\mch$ should be smaller than 620 GeV.
Since the $\tb$ dependence on the upper bounds is negligible as shown in 
the left panel of Fig.~\ref{fig-theoretical},
the presence of
the upper bounds on $\mhh$, $m$, and $\mch$ is a generic feature of the 2HDM-SM4. 
This result is unexpected.
In the so-called normal mass hierarchy scenario where the observed 125 GeV scalar is the lighter $h$,
it is usually expected that the theory can hide in the decoupling region.
However the exact wrong-sign limit does not allow decoupling of the theory.
This is a unique feature of this model.

\section{Indirect Constraints from the oblique parameters and the Higgs precision data}
\label{sec:indirect}

In this section
we narrow down the parameter space
by applying the indirect constraints from the electroweak oblique parameters and the Higgs precision data.

\subsection{Electroweak oblique parameters $\Dt S$ and $\Dt T$}

The oblique parameters $\Dt S$ and $\Dt T$
get affected by the sequential fourth generation fermions~\cite{Kribs:2007nz,Dighe:2012dz}
as well as by new scalar bosons~\cite{Grimus:2007if,Grimus:2008nb}.
The current experimental data are consistent with the SM values, given by~\cite{Patrignani:2016xqp}
\bea
\label{eq:PDG:ST}
\Dt S = 0.05 \pm 0.10,\quad
\Dt T = 0.08 \pm 0.12,
\eea
where the parameter $\Dt T$ is more sensitive to new particles.
It is well known that new contributions to $\Dt T$
are suppressed when the new particles running in the self-energy diagrams of gauge bosons
have the same masses.
We require that $\mtp \simeq \mbp$, $\mtaup \simeq \mnup$, and
two masses among $\mhh$, $\ma$, and $\mch$ are degenerate.
We also note that
the contributions from $H$ and $A$ to the oblique parameter $\Dt T$
are negative~\cite{Branco:2011iw} while those from the fourth generation fermions
are positive~\cite{Kribs:2007nz,Dighe:2012dz,Grimus:2008nb}.
Large mass splittings in the scalar and fourth generation fermion sectors
are allowed if exquisite cancellation occurs.
In this work, however, we do not consider the conspiracy between the new scalar sector and the new fermion sector
in explaining the oblique parameters.

\subsection{Higgs precision data}
We take the combined analysis of ATLAS and CMS  on the Higgs coupling modifier $\kp_i$
based on the LHC Run 1 data,
corresponding
to integrated luminosities per experiment of $5\ifb$ at $\sqs=7\tev$
and $20\ifb$ at $\sqs=8\tev$~\cite{Khachatryan:2016vau}\footnote{
More recent analyses
of the 13 TeV data~\cite{Aaboud:2017vzb,Sirunyan:2017exp} are neither
combined ones of the ATLAS and CMS nor suitable for $\kp_t=-\kp_b$.}.
The analysis
is based on a few assumptions,
each of which constrains $\kp_i$'s differently.
The 2HDM-SM4 model belongs to the category where new loop couplings beyond the SM (BSM) are allowed.
The analysis result is that all of $\kp_i$'s are consistent with the SM value.
For the allowed values of $\kp_i$'s in this category,
we refer the reader to Fig.~15 in Ref.~\cite{Khachatryan:2016vau}.
%

\begin{figure}[t!]
\begin{center}
\includegraphics[width=0.5\textwidth]{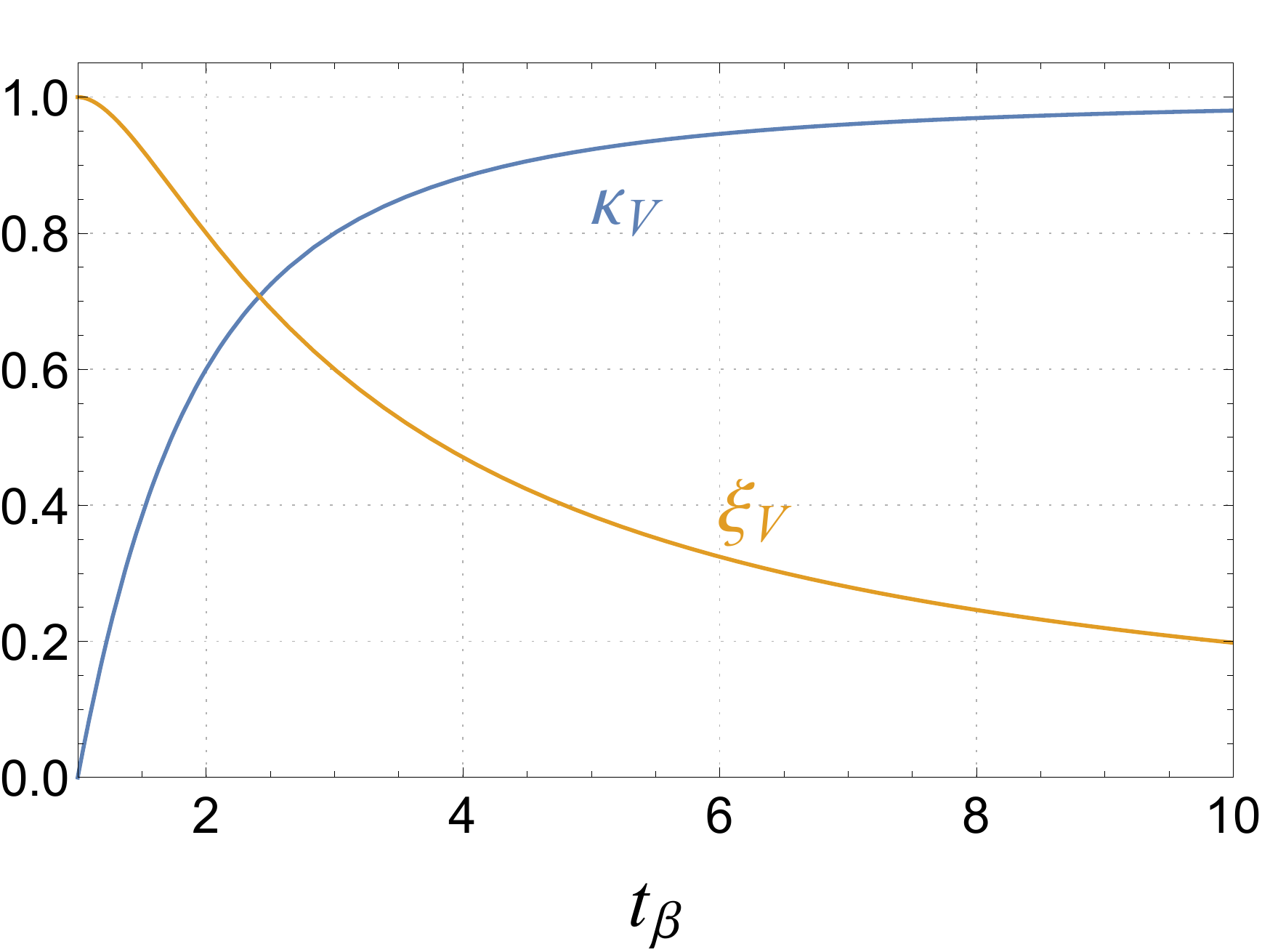}
\caption{\label{fig-kappaVxiV}
\baselineskip 3.0ex
$\kp_V$ and $\xi_V$ as a function of $\tb$.
}
\end{center}
\end{figure}

The exact wrong-sign limit naturally explains the SM-like $\kp_t$, $|\kp_{b,\tau}|$, $\kp_g$ and $\kp_\gm$.
One minor concern is $ \left| \kp_b \right|$,
of which the maximum value at $2 \sigma$ level is about 5\% smaller than one.
Since a small deviation from the exact wrong-sign limit can easily accommodate this result,
we stick to the exact wrong-sign limit as our reference point.
On the other hand,
$\kp_V$ can significantly deviate from one.
In Fig.~\ref{fig-kappaVxiV},
we show $\kp_V$ and $\xi_V$ as a function of $\tb$.
The observed $\kp_V$ requires very large $\tb$:
if $\kp_V  =0.98$, we need $\tb\approx 9.95 $.
However too large $\tb$
violates the perturbativity of the Yukawa couplings of $b'$ and $\tau'$
with $H$ and $A$, $Y^{H/A}_{b',\tau'}=\tb M_{b',\tau'}/v$.
If $\kp_V  =0.98$ and $\mbp = 340\gev$, for example,
we have
 $Y^H_{b'} = 13.8$.
In summary, there is a tension on the value of $\tb$:
the observed $\kp_V$ pushes $\tb$ upward;
the perturbativity of the $H \bbp$ coupling presses $\tb$ downward.

\begin{figure}[t!]
\begin{center}
\includegraphics[width=0.55\textwidth]{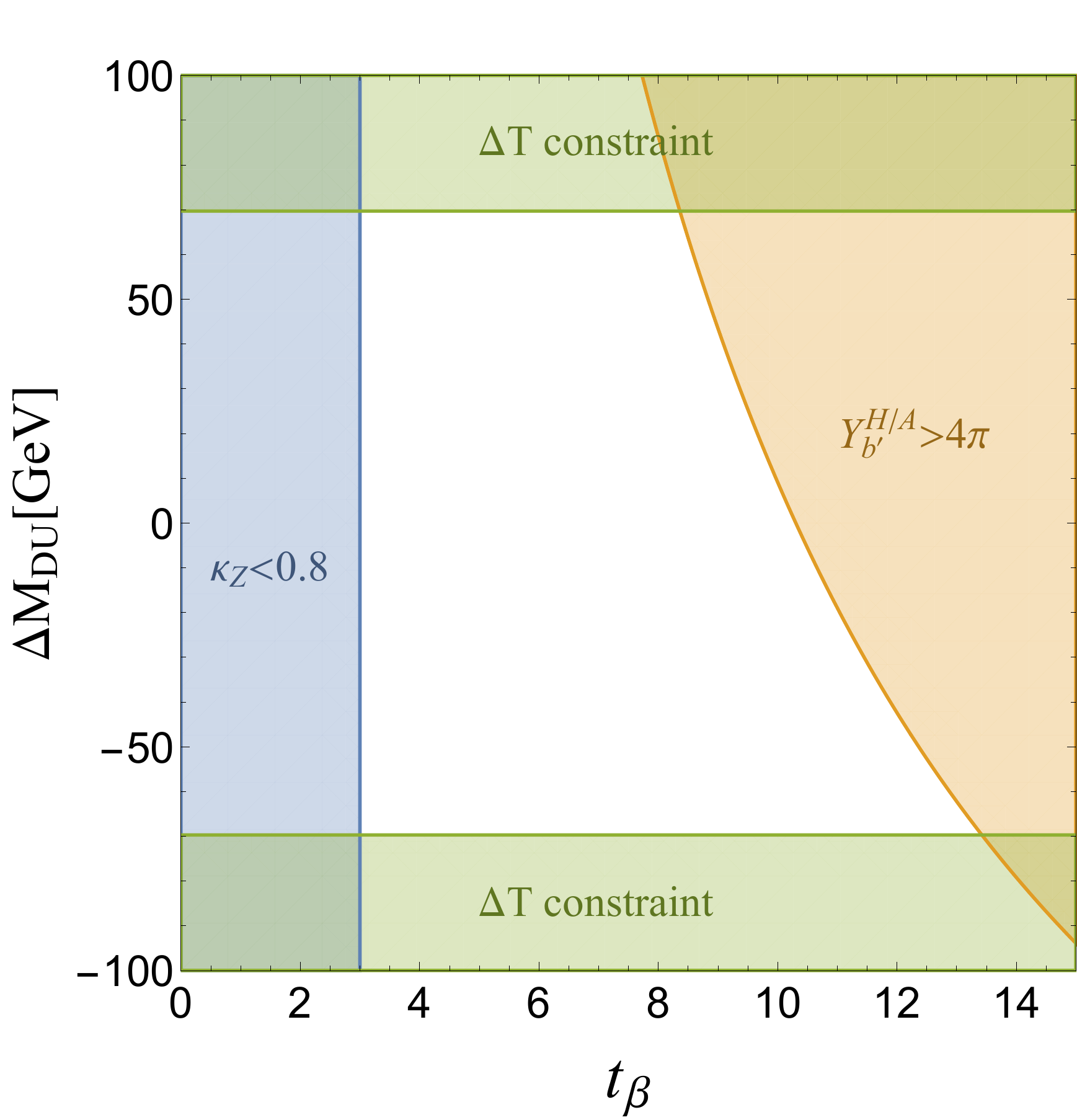}
\caption{\label{fig-constraints}
\baselineskip 3.0ex
The excluded regions of $(\tb,\Dt M_{DU})$
by the observed Higgs coupling modifier $\kp_V$ at $2\sigma$ level,
the oblique parameter $\Dt T$ at $2\sigma$ level, and
the perturbativity of $Y^{H/A}_{b'}$.
Here $\Dt M_{DU} = \mbp - \mtp=\mtaup - \mnup$ for the fixed $M_U =300\gev$.
For the oblique $\Dt T$ parameter, we set $\mhh=\ma=\mch=600\gev$.
}
\end{center}
\end{figure}

In Fig.~\ref{fig-constraints}, we summarize the constraints from the oblique parameter $\Dt T$  at $2\sigma$ level,
 the Higgs coupling modifier $\kp_V$  at $2\sigma$ level, and
the perturbativity of $Y^{H/A}_{b'}$.
Here $\Dt M_{DU} = \mbp-\mtp=\mtaup-\mnup $ and $M_U(=\mtp=\mnup)$ is fixed to be 300 GeV.
The observed $\kp_V$
allows $\tb \gsim 3$,
while the perturbativity of $Y^{H/A}_{b',\tau'}$
requires $\tb \lsim 8$.
The oblique parameter $\Dt T$ demands that the mass of an up-type fourth generation fermion
be similar to that of the corresponding down-type fermion like $\Dt M /M \lsim 20\%$.

In addition,
the experimental result on exotic Higgs decay $\br_{\rm BSM}^h \leq 0.34$~\cite{Khachatryan:2016vau}
has an important implication on the mass of the pseudoscalar $A$.
In the 2HDM, $\ma$ is a free parameter so that
$A$ can be light enough to kinematically allow
$h \to AA$.
The observed oblique parameter $\Dt T$ can be explained by the mass degeneracy 
$\mhh \simeq \mch$.
When writing $\lg \supset  \lm_{hAA} A A h /2$,
the partial decay rate is
\bea
\Gm(h \to AA) = \frac{1}{32\pi  \mh} \lm_{hAA}^2 \sqrt{1 - \frac{4 \ma^2}{\mh^2}}.
\eea
In the general 2HDM, $\lm_{hAA}$ is
unknown for the given $\mh$ and $\ma$
because of an additional free parameter $m^2$.
In the exact wrong-sign limit ($\al+\beta=\pi/2$), however, the $m^2$ term is proportional to $c_{\beta+\al}$
and thus vanishes. The value of $\lm_{hAA}$ is determined by $\tb$ and $\ma$ as
\bea
\label{eq:lm:hAA}
\lm_{hAA} = \frac{c_{2\bt}}{v}
\left[
2 \ma^2 - \mh^2
\right].
\eea
The observed $\br^h_{\rm BSM} \leq 0.34$ is translated into $\Gm(h \to AA) \lsim  \Gm^{h,\sm}_\tot/2$,
which can be satisfied when $ 0.9 \lsim \tb \lsim 1.1$ ($ 0.62 \lsim \tb \lsim 1.60$)
for $\ma \ll \mh$ ($\ma=62\gev$).
This small $\tb$ region is excluded
by the observed $\kp_V$: see Fig.~\ref{fig-kappaVxiV}.
In summary, a light pseudoscalar boson with $\ma \lsim \mh/2$ in the 2HDM-SM4
is excluded by the observed $\br_{\rm BSM}^h$.

\section{Decay and Production of $H$ and $A$}
\label{sec:H:A}

\subsection{Decays}
\label{subsec:decay}
In this section, we discuss
the decay and production of neutral heavy Higgs bosons, $H$ and $A$.
Considering the theoretical and direct search bounds on $M_F$,
the perturbative unitarity of the scalar-scalar scattering,
and the $b \to s \gm$ constraint altogether,
we take the following benchmark scenario:
\bea
\label{eq:benchmark}
m&=&\mhh, \quad
\ma \approx \mhh, \quad \mch=580\gev,
\\ \nn
\mtp&=&300\gev,\quad \mbp=340\gev,\quad \mnup=430\gev ,\quad \mtaup=380\gev.
\eea
Kinematically, the decays of $H \to AA$ and $H \to H^+ H^-$ are prohibited.
Note that the decays of 
$H \to W^\pm H^\mp$ and $A \to Zh$ are possible
through the interaction Lagrangian of
\bea
\label{eq:Lg:Vss}
\lg \supset
\frac{g_Z}{2} \cba Z_\mu h \stackrel{\leftrightarrow}{\rd^\mu} \!\! A +
i  \frac{g }{2} \sba \, \left[  W_\mu   H \stackrel{\leftrightarrow}{\rd^\mu} \!\!  H^-   + H.c. \right]
\,,
\eea
where $g_Z = g/c_{\theta_W}$, $\theta_W$ is the weak mixing angle,
and $\phi \stackrel{\leftrightarrow}{\rd^\mu}\!\! \eta = \phi \,\rd^\mu \eta - (\rd^\mu \phi)\eta$.

\begin{figure}[h]
\begin{center}
\includegraphics[width=.6\textwidth]{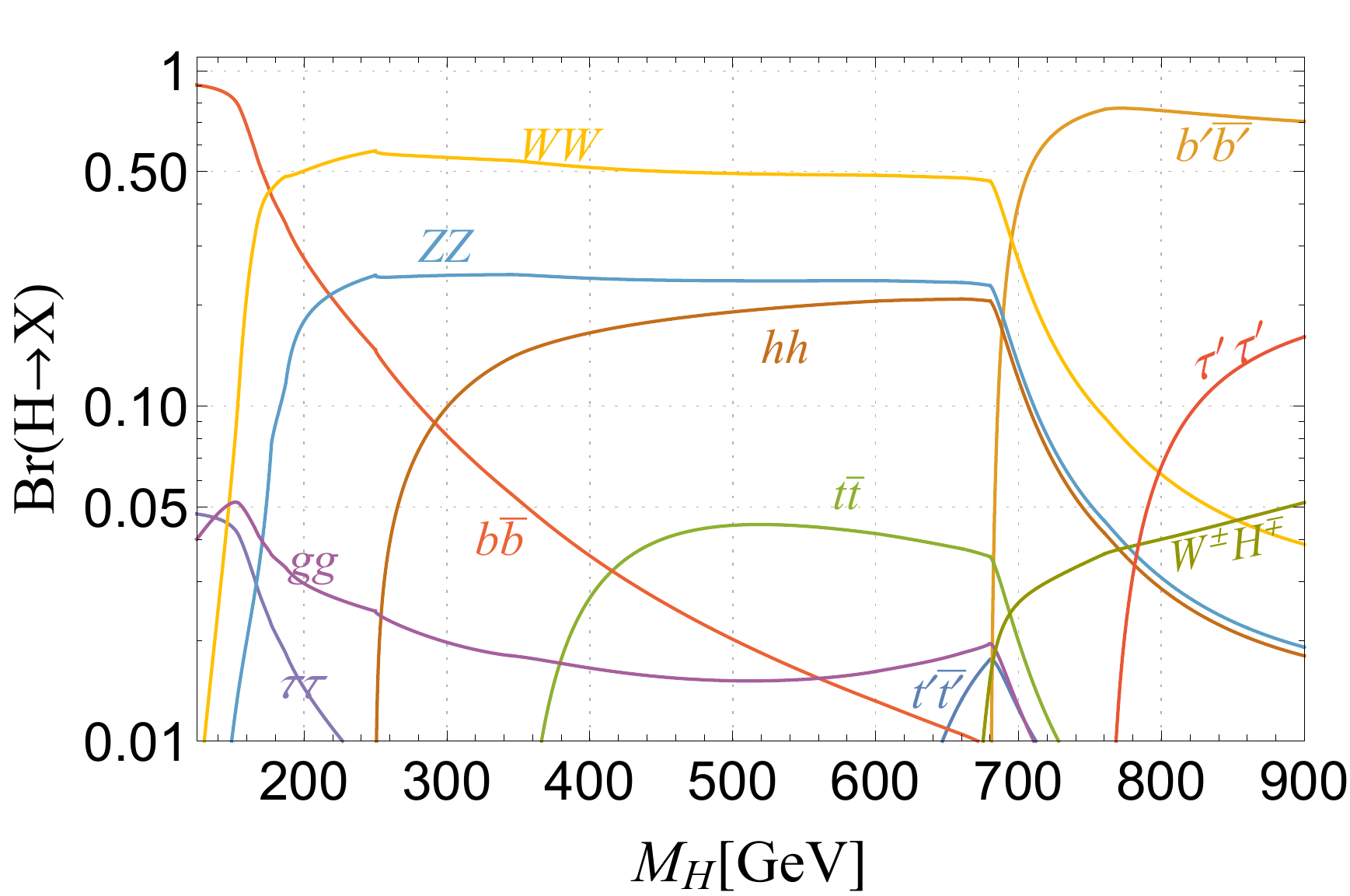}
\end{center}
\caption{\label{fig-BrH}
The branching ratios of $H$
as a function of $\mhh$ for $\tb=4$.
We set
$\mtp=300\gev$, $\mbp=340\gev$, $\mnup=430\gev$, $\mtaup=380\gev$, $m=\mhh$,
$\ma = \mhh$, and $\mch=580\gev$.
}
\end{figure}

In Fig.~\ref{fig-BrH}, we show the branching ratios of $H$
as a function of $\mhh$ for $\tb=4$ and $m=\mhh$.
Before the $W W$ threshold,
the dominant decay mode of $H$
is into $\bb$,
followed by $\ttau$ and $gg$ modes
because of the $\tb$ enhancement in $Y^H_{b,\tau}$.
After the $W W/Z Z$ threshold, non-zero $\xi_V=\cba$ ($\cba \approx 0.47$  for $\tb=4$)
yields dominant decay of $H$ into $W W$ and $Z Z$, 
because of 
the longitudinal polarization enhancement in the heavy scalar decay
into a massive gauge boson pair~\cite{Yoon:2017wul}.

For $\mhh> 2 \mh$,
the decay of $H \to h h$ becomes also important.
The triple Higgs coupling $\lm_{Hhh}$ in the exact wrong-sign limit is
\bea
\label{eq:lm:Hhh}
\lm_{Hhh} =
\frac{\cba}{v}
\lf
m^2 - \frac{1}{2} \mhh^2 - \mh^2
\ri
.
\eea
With the
sizable $\cba$ 
and the condition of $m \simeq \mhh$ from the theoretical constraints shown in Fig.~\ref{fig-theoretical},
the branching ratio of $H\to h h$ is substantial.
Above the $\ttop$ threshold,
$H \to \ttop$ mode turns on, but not dominantly because
$H \ttop$ coupling is inversely proportional to $\tb$.
For the same reason, $H \to t'\bar{t}'$ mode is also minor even after the $\ttopp$ threshold.
After the $\bbp$ and $\ttaup$ threshold, $H \to b'\bar{b}'$
is dominant and $H \to \ttaup$ is second dominant.
The third dominant decay channel is into $H \to W^\pm H^\mp$,
which always remains important because its vertex is proportional to $\sba$: see Eq.~(\ref{eq:Lg:Vss}).

\begin{figure}[h]
\begin{center}
\includegraphics[width=.6\textwidth]{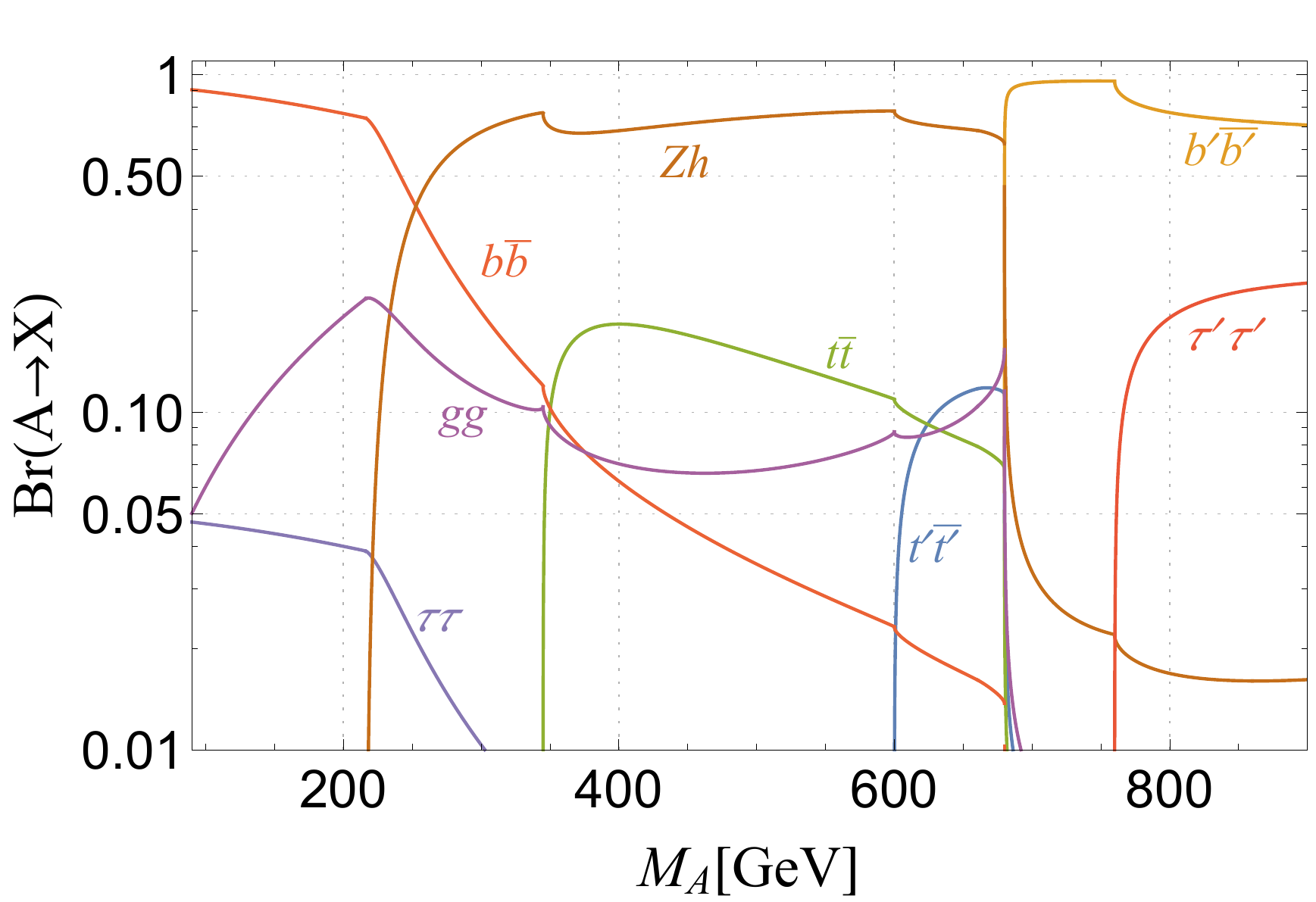}
\end{center}
\caption{\label{fig-BrAtb4}
\baselineskip 3.0ex
The branching ratios of $A$
as a function of $M_{A}$ for $\tb=4$.
We set
$\mtp=300\gev$, $\mbp=340\gev$, $\mnup=430\gev$, $\mtaup=380\gev$, $m=\mhh$,
$\ma = \mhh$, and $\mch=580\gev$.
}
\end{figure}

The branching ratios of the CP-odd Higgs boson $A$
as a function of $\ma$ are shown in Fig.~\ref{fig-BrAtb4}.
We take the benchmark scenario in Eq.~(\ref{eq:benchmark}).
Because of the CP-odd nature,
$A$ does not decay into $W W$, $Z Z$, or $hh$,
but does decay into $Zh$ since 
the $AZ h$ vertex is proportional to non-negligible $\cba$.
Before the $Zh$ threshold,
$\bb$ mode is dominant:
the $A\bb$ coupling also has the $\tb$ enhancement.
The second dominant decay mode into $gg$
has much larger branching ratio than the $\ttau$ mode,
unlike the case of $H$.
It is because of the larger
$gg$ loop function for the CP-odd scalar than that for the CP-even scalar.
When $ m_Z+\mh < \ma < 2 \mbp$,
the decay of $A \to Z h$ is dominant,
contrary to the alignment limit.
In this mass range, the decay into $gg$ is still important.
We expect a sizable cross section of gluon fusion production of $A$.
The branching ratios of the $\ttop$ and $\ttopp$ modes are not as large as that of the $Zh$ mode
because both $A\ttop$ and $A\ttopp$ couplings are inversely proportional to $\tb$.
After the $b'\bar{b}'$ and $ \ttaup$ threshold,
$A$ decays into $\bbp$ dominantly, followed by the $\ttaup$ mode.
The next dominant mode is into $Zh$ with
$\br(A \to Zh) \sim  \mco(1)\%$.

\begin{figure}[t!]
\begin{center}
\includegraphics[width=.65\textwidth]{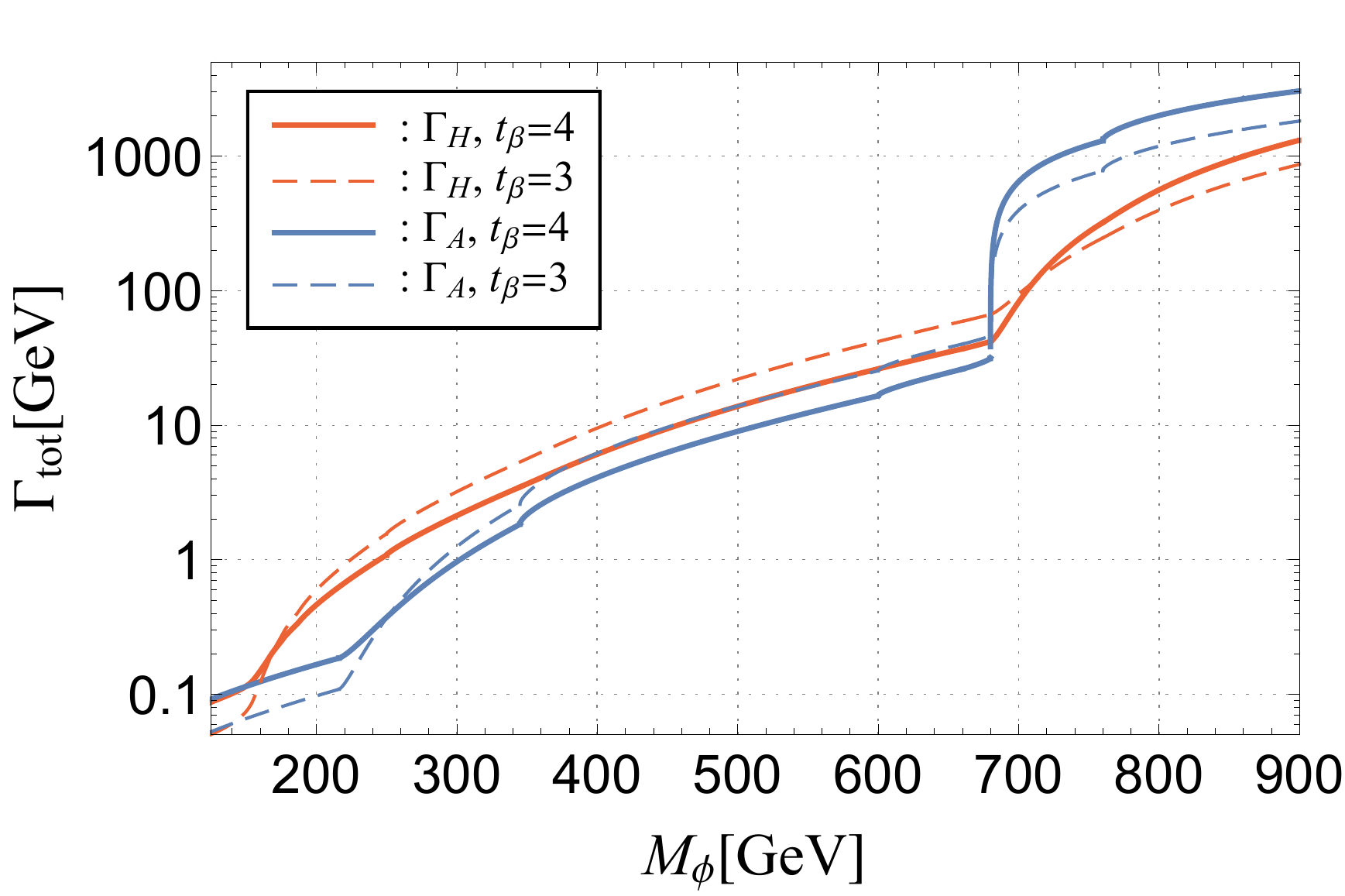}
\end{center}
\caption{\label{fig-gammatot}
\baselineskip 3.0ex
Total decay width of $H$ and $A$ as a function of their masses
for two $\tb$ cases, $\tb=3$ and $\tb=4$.
We set $m=\mhh$,
$\mtp=300\gev$, $\mbp=340\gev$, $\mnup=430\gev$, and $\mtaup=380\gev$,
and do not include the decays of $H \to H^+ H^-$ and $H \to AA$.
}
\end{figure}

In Fig.~\ref{fig-gammatot},
we show the total decay widths of $H$ and $A$ as a function of $\mhh$ and $\ma$
for $\tb=3$ and $\tb=4$ in the benchmark scenario.
For both $H$ and $A$,
$\Gm_\tot$ increases rapidly with its mass,
especially above the $\bbp$ threshold.
The dependence of $\Gm_\tot^{H/A}$ on $\tb$ is interesting.
Before the $\bbp$ threshold,
the larger $\tb$ is, the smaller the total decay width is.
This behavior is due to the fact that
the major decay modes before the $\bbp$ threshold,
$H \to W W/Z Z$ and $A\to Zh$,
have the partial decay rates all proportional to $\xi_V^2 $ which decreases with increasing $\tb$:
see Fig.~\ref{fig-kappaVxiV}.
After the $\bbp$ threshold, on the other hand, 
the dominant decay rate
$\Gm(H/A \to \bbp)$ is proportional to $\tb^2$.

Another crucial question related to $\Gm_\tot$
is whether $H$ and $A$ can be probed as a narrow resonance
so that the number of the new events is proportional to the production cross section
times the branching ratio.
For reference, the LHC criteria as a narrow resonances is $\Gm/M \leq 1\%$
in the $\rr$ channel \cite{Aaboud:2018xdt,CMS:2017rli},
$\Gm/M \leq 0.5\%$ in the $Z Z$ channel~\cite{Aaboud:2017vzb,Chatrchyan:2013mxa},
and $\Gm/M \leq 15\%$ in the dijet channel~\cite{Aaboud:2017yvp}.
When $\tb=4$,
for example,
heavy Higgs bosons like $\mhh\gsim 350\gev$ and $\ma \gsim 300\gev$
do not belong to the narrow width category, which requires
going beyond $\sg \times \br$.
Particularly above the $\bbp$ threshold,
both $\Gm_\tot^H$ and $\Gm_\tot^A$
are so large like $\Gm_\tot \sim \mhh$.
It is almost impossible
to observe a mass peak in this mass region.
We need a new strategy.


%
In order to deal with the very large width case like $\Gm_\tot^{H/A} \sim M_{H/A}$,
two points should be considered.
First,
we need full calculation of $\sg(pp \to ij)$ including the SM continuum background 
in order not to miss the significant interference.
%
The second point is
that new events  spread out over multiple $m_{ij}$ bins,
not only to the bin of $m_{ij}=M$.
Without the possibility of observing a mass peak,
we may rely on counting total events, which requires a very good understanding of the background.
Or we can utilize the results of usual analysis of the resonance searches
for the excess in the invariant mass bins,
which takes the following steps:
(i) events are collected in a specific $m_{ij}$ bin; 
(ii) the number of the events in the bin is compared with that of the expected background;
(iii)
no excess leads to the upper bounds on $\sg \times \br(X \to ij)$ for a possible new particle with mass $m_{ij}$.
Since multiple $m_{ij}$ bins get affected by a single new particle when $\Gm \sim M$,
we need to calculate the excess of each $m_{ij}$ bin nearby $M$
and  compare with the upper bounds on $\sigma \times \br (A \to Zh)$ for the corresponding bin.
The size of each $m_{ij}$ bin, $\Dt M$, depends on the experimental analysis.
Since our $d \sg^\np$ already includes the SM contributions,
the excess over the SM backgrounds corresponds to the difference between the full
$d \sg^\np$ and $d \sg^\sm$.
In order to compared with the excess in the $m_{ij}\in [M,M+\Dt M]$ bin,
therefore,
we calculate the partially integrated cross section $\Dt \sg^\np $ given by
\bea
\label{eq:Dsgnp}
\Dt \sg^\np =
\int_{M}^{M+\Dt M} d m_{ij}
\lfb
\frac{d \sg^\np(gg\to ij)}{d m_{ij}} - \frac{d \sg^\sm(gg\to ij)}{d m_{ij}}\rib
.
\eea
We shall use this method when constraining very heavy $A$ and $H$ in the $Z h$ and $Z Z$ final states, respectively.

\subsection{Productions}

\begin{figure}[h!]
\begin{center}
\includegraphics[width=.7\textwidth]{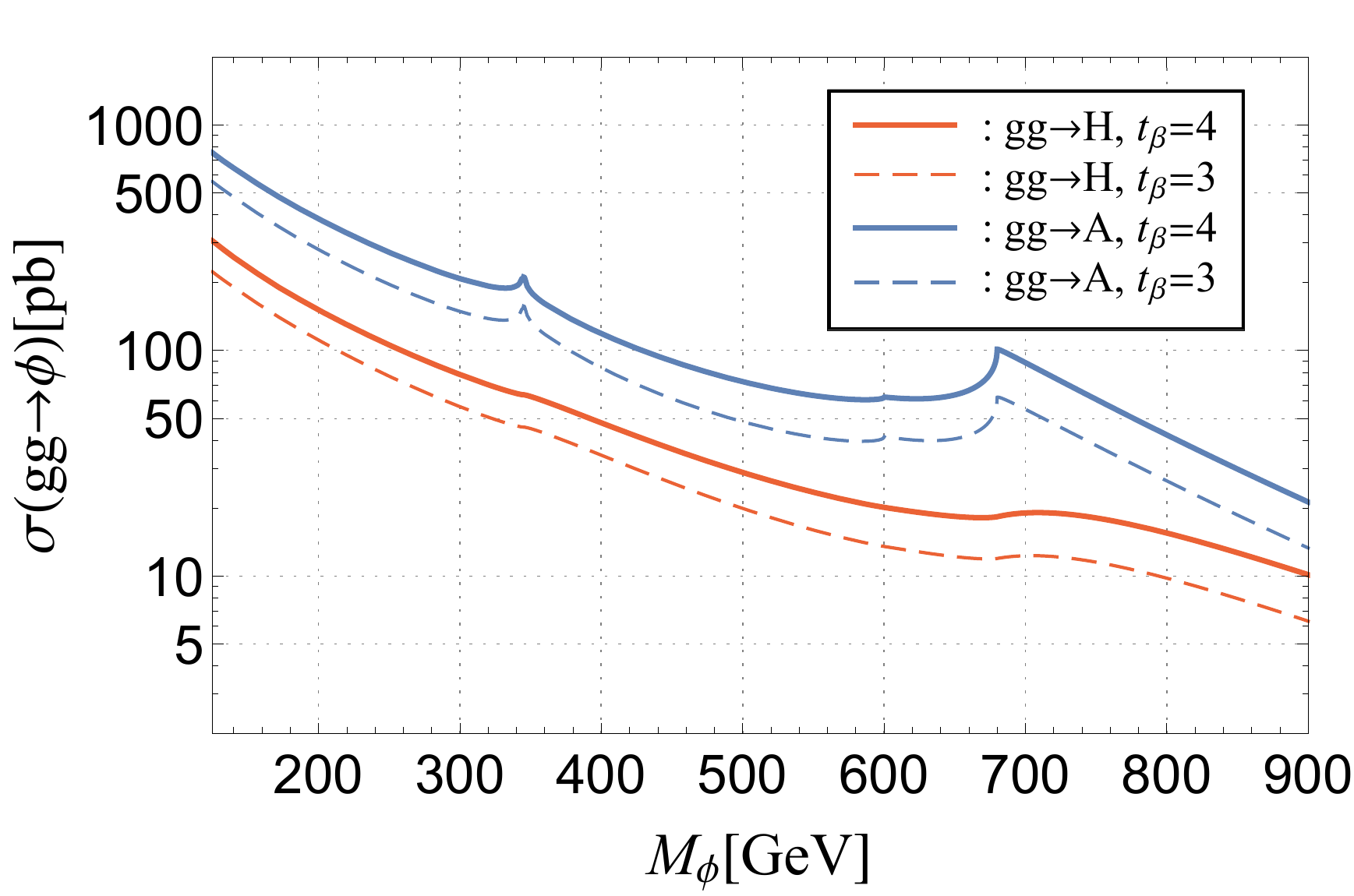}
\end{center}
\caption{\label{fig-prodtot}
\baselineskip 3.0ex
The gluon fusion production cross sections of $H$ and $A$
as functions of $\mhh$ and $\ma$ at the 13 TeV LHC
for $\tb=3$ and $\tb=4$.
We apply NNLO $K$ factor to the heavy Higgs resonance production by using \texttt{HIGLU} package~\cite{Spira:1995mt}.
}
\end{figure}

We study the production of $H$ and $A$.
The production cross section of $H/A$ with a small width
is~\cite{Franceschini:2015kwy}
\bea
\label{eq:proXS}
\sg (pp \to H/A ) = \frac{1}{s \, M_{H/A}}
\sum_{\mathfrak{p}} C_{\mathfrak{p p}} \Gm(H/A \to \mathfrak{p p}) ,
\eea
where $\mathfrak{p}$ is a parton in a proton
and $C_{\mathfrak{p p}}$ is the $\mathfrak{p p}$ dimensionless partonic integral.
Because $C_\bb/ C_{gg} \sim 0.01$ at the 13 TeV LHC
and moderate $\tb \in [3,8]$ implies not too large $H\bb$ and $A\bb$ couplings,
we consider only the gluon fusion production.
Figure \ref{fig-prodtot}
presents the gluon fusion production cross sections of $H$ and $A$ at the 13 TeV LHC
as a function of $\mhh$ and $\ma$ respectively.
We consider $\tb=3$ and $\tb=4$,
and include the 
NNLO $K$ factor  by using \texttt{HIGLU} package~\cite{Spira:1995mt}.
In the whole mass ranges,
the production cross section is large:
for $\mhh,\ma=600\gev$, for example, $\sg (pp \to H/A ) \sim \mco(10) \pb$.
The pseudoscalar $A$ has larger cross section than $H$,
because of larger loop function for $gg$.
As $\tb$ increases,
$\sg(gg \to H/A)$ increases because
the contribution from the $b'$ quark in the loop
is enhanced.
For the CP-odd $A$, the threshold effects at $\ma=2m_t$ and $\ma =2 \mbp$
are more prominent due to the cusp structure of the real part of the $gg$ form factor $A^A_{1/2}(\tau)$~\cite{Djouadi:2005gj}.
We caution the reader that the production cross section in Eq.~(\ref{eq:proXS})
for $M_{H/A} \gsim 2 \mbp$ gives just a rough estimation.
When $\Gm_\tot \sim M_{H/A}$, the production of $H/A$ itself
is not meaningful.
We need to set
the final states $ij$
and to perform
the full calculation of $\sg(pp \to ij)$ including the SM continuum background because
the interference effects crucially depend on 
the final state.

\section{Constraints from the direct searches}
\label{sec:direct:search}

In this section, we study the constraints from the direct searches
for neutral scalar bosons at high energy colliders.
The allowed region for the charged Higgs boson mass,
$570 \lsim \mch \lsim 620\gev$, is very difficult
to probe because of extremely small production rate of $H^\pm$~\cite{Miller:1999bm}.
Our main target channels are summarized in Table \ref{tab:search:HA}.
Brief comments on $\rr$, $\mmu$, and $\bb$ modes are in order here.
Although the very clean $\rr$ mode is searched for from $m_{\rr}=65\;(70)\gev$
by the ATLAS (CMS) experiments~\cite{Aad:2014ioa,CMS:2017yta},
it does not constrain the model since
the branching ratio is very small: $\br(A\to \rr ) \sim \mco(10^{-5})$ for $\ma=65\gev$.
The data in the $\mmu$ channel~\cite{Aad:2016naf}
are also insufficient yet because of the extremely small $\br(H/A \to \mmu) \lsim \mco \lf 10^{-4} \ri$.
The $\bb$ mode constrains a NP model
only in the heavy mass range of $m_{\bb}\geq 800\gev$~\cite{Aaboud:2018tqo}
due to
the huge QCD background,
where
both $\br(H \to\bb)$ and $\br(A\to\bb)$
are extremely small because of the dominant decays into $\bbp$.

\begin{table}[h!]
\begin{center}
{\renewcommand{\arraystretch}{1.2}
\begin{tabular}{|c|c|c|c|c|c|c|c|c|}
\hline
process & target & mass range & experiment \\ \hline
~~~$\ee\to 4b,4\tau,\bb\ttau$~~~ & $A$ & $[2 m_\tau, 100\gev]$ & LEP~\cite{Schael:2006cr} \\ \hline
\multirow{2}{*}{$pp \to \ttau$} & \multirow{2}{*}{$H$, $A$} & $[100\gev,1\tev]$ & LHC Run 1~\cite{Aad:2014vgg,Khachatryan:2014wca} \\
 &  & $[90\gev,3.2\tev]$ & LHC Run 2~\cite{Aaboud:2017sjh,Sirunyan:2018zut}
 \\ \hline
$pp \to Z Z^{(*)}$ & $H$  & $[110\gev,1\tev]$ & LHC Run-2~\cite{Sirunyan:2018qlb,Aaboud:2017itg,Khachatryan:2015cwa,Aad:2015kna}\\ \hline
$pp \to Z h$ & $A$  & $[200,1000]$ & LHC Run-1~\cite{Khachatryan:2015lba,Aad:2015wra}\\ \hline
\end{tabular}
}
\caption{\label{tab:search:HA}
Summary of the direct searches efficient for $H/A$ in the 2HDM-SM4 at high energy colliders.}
\end{center}
\end{table}
\subsection{Constraints from the direct searches on $A$}

Meaningful constraint is
from the neutral Higgs boson searches at the LEP
through $\ee \to H_i H_j$ $(H_{i,j}=h,H,A)$
in the framework of the Minimal Supersymmetric Standard Model~\cite{Schael:2006cr}.
The analysis was based on four different decay channels of $4b$, $2b2\tau$, and $4\tau$.
Since the searches span a center of mass energies from $91\gev$ to $209\gev$,
the heavy CP-even $H$ cannot be produced kinematically.
The CP-odd scalar boson $A$ is produced in association with $h$,
mediated by the $Z$ boson.
With the observed Higgs boson mass of 125 GeV, the LEP result excludes $\ma \lsim 65\gev$.
This is consistent with the exclusion from $\br^h_{\rm BSM}$,
as discussed below Eq.~(\ref{eq:lm:hAA}).

\begin{figure}[h!]
\begin{center}
\includegraphics[width=.6\textwidth]{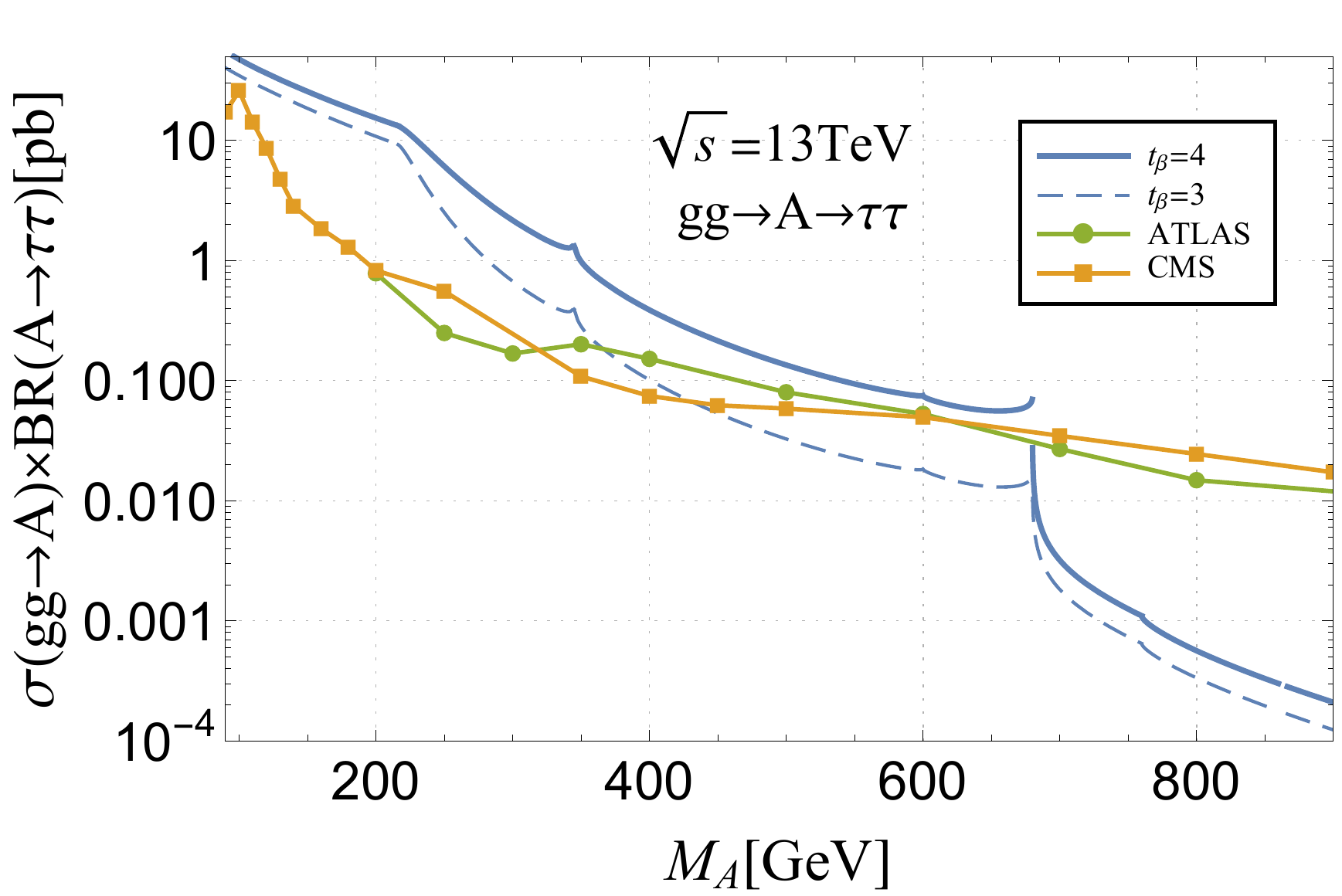}
\end{center}
\caption{\label{fig-Atautau13}
\baselineskip 3.0ex
$\sg(gg\to A)\times\br(A \to \ttau)$
as a function of $\ma$ at the 13 TeV LHC for $\tb=3$ and $\tb=4$,
We apply NNLO $K$ factor to the heavy Higgs resonance production by using \texttt{HIGLU} package~\cite{Spira:1995mt}.
For comparison,
we also show the  95\% C.L. upper limits on $\sg \times \br (\phi \to \ttau)$ data~\cite{Sirunyan:2018zut,Aaboud:2017sjh}.
}
\end{figure}

We now consider the resonance searches in the $\ttau$ channel at the LHC.
Both ATLAS and CMS experiments
presented their results
based on the Run-1~\cite{Aad:2014vgg,Khachatryan:2014wca} and
Run-2~\cite{Sirunyan:2018zut,Aaboud:2017sjh} data.
Since the LHC Run-1 data at $\sqrt{s}=7+8\tev$
put weaker constraints, we focus on the Run-2 results.
In Fig.~\ref{fig-Atautau13},
we show $\sg\times\br(A\to\ttau)$ in the 2HDM-SM4 at the 13 TeV LHC.
As shall be shown in the next sub-section,
the pseudoscalar $A$ gets much stronger constraint from the $\ttau$ channel than the CP-even $H$.
It is partly because the gluon fusion production of $A$ is more efficient
due to larger loop function than that of $H$:
see Fig.~\ref{fig-prodtot}.
Another reason is larger branching ratio of $A\to\ttau$
because of the absence of $A \to W W/Z Z/hh$.
Both ATLAS and CMS experiments exclude the parameter space with $\tb \geq 4$ and $\ma \leq 2 \mbp$.
When $\tb=3$ (the minimum value of $\tb$ allowed by the observed $\kp_V$),
two experiments yield different lower bounds on $\ma$.
Upon the absence of a combined ATLAS and CMS analysis yet,
we take a conservative stance on constraining the model, i.e.,
adopting the weaker constraint between two experiment results.
For $\tb=3$,
$\ma \lsim 350 \gev$ is excluded at the 95\% C.L.

\begin{figure}[h]
\begin{center}
\includegraphics[width=.8\textwidth]{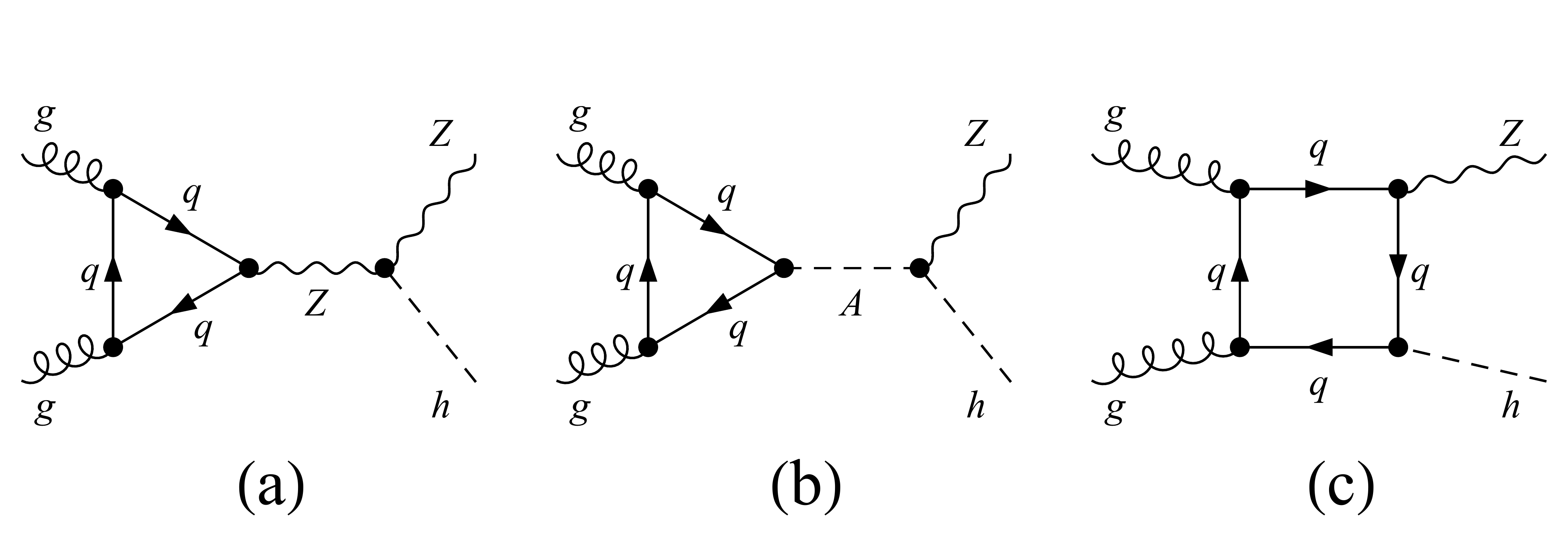}
\end{center}
\caption{\label{fig-Feynman-Zh}
\baselineskip 3.0ex
Feynman diagrams for the $gg\to Z h$ process in the 2HDM-SM4.
Here $q$ denotes all of the four generation quarks, including $t'$ and $b'$.
}
\end{figure}

The smoking-gun signature for $A$ at the LHC is the $Zh$ channel,
followed by $Z \to \ell\ell$ and $h\to \bb/\ttau$~\cite{Aad:2015wra,TheATLAScollaboration:2016loc,CMS:2018xvc}.
In the SM, $pp \to Zh$ proceeds mainly through $\qq \to Z h$, mediated by $Z^*$.
The gluon fusion production also occurs in the SM
through the quark triangle diagram  and the quark box diagram,
(a) and (c)
in Fig.~\ref{fig-Feynman-Zh}.
In the SM, $\sg(gg \to Zh)$
is small, about 10\% of $\qq$ annihilation process,
mainly because of the  \emph{destructive}
interference between the triangle and box diagrams~\cite{Englert:2013vua}.
In the 2HDM-SM4,
there are four kinds of
new contributions:
(i) $\sg(\qq \to Zh)$ is reduced by the factor of $\kp_V^2$;
(ii) $\br(h \to \bb/\ttau)$ in the exact wrong-sign limit is increased 
because of smaller $\Gm(h\to W W/Z Z)$ than in the SM
but the same
$\Gm (h \to \bb/\ttau)$;
(iii) the fourth generation quarks contribute to all of the loop diagrams for $gg \to Zh$;
(iv) new triangle diagrams mediated by $A$ appear.
In what follows,
we call Fig.~\ref{fig-Feynman-Zh}(a) the $Z$-triangle diagram,
Fig.~\ref{fig-Feynman-Zh}(b) the $A$-triangle diagram,
and Fig.~\ref{fig-Feynman-Zh}(c) the box diagram.

\begin{figure}[t]
\begin{center}
\includegraphics[width=0.7\textwidth]{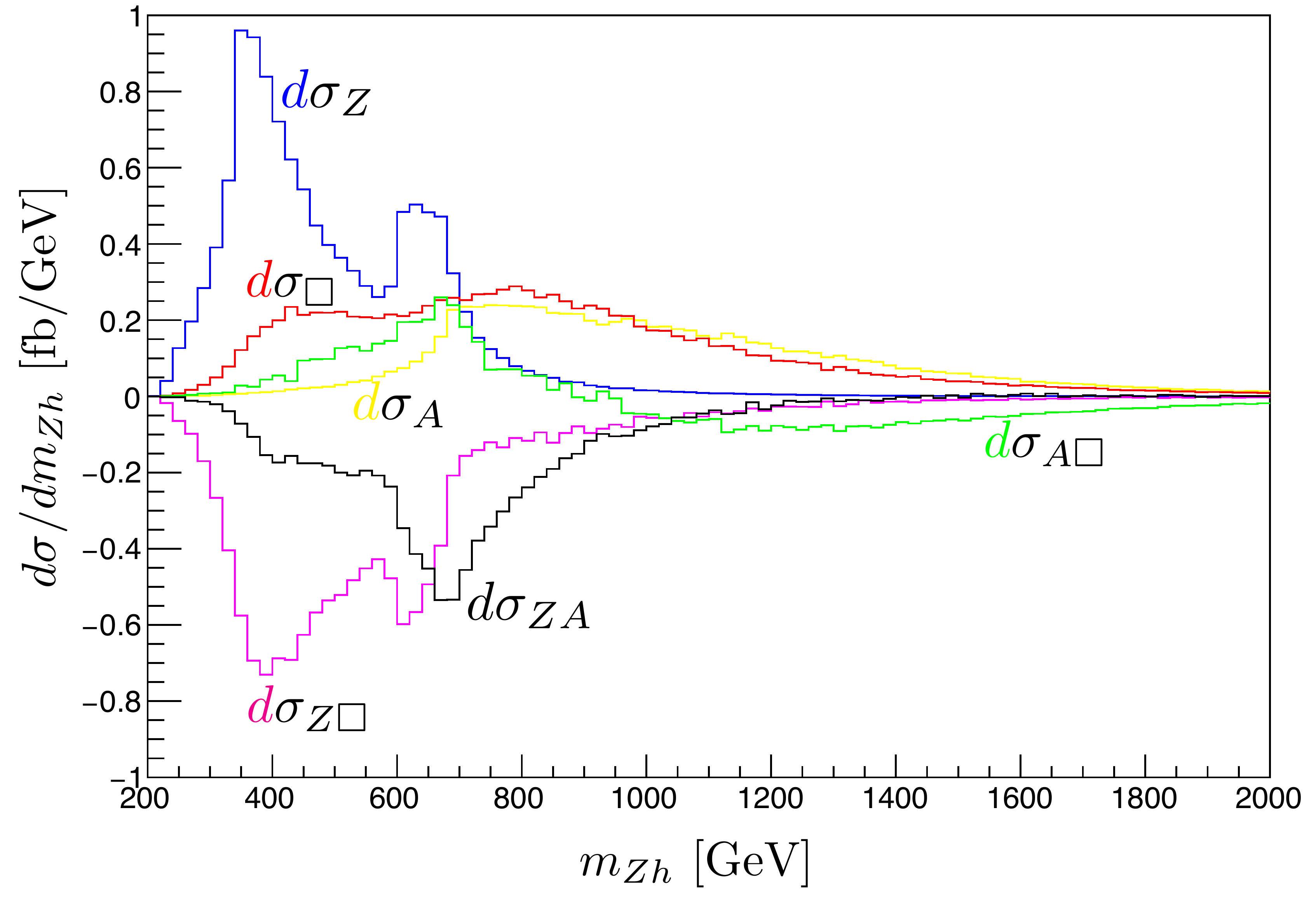}
\end{center}
\caption{\label{fig-XSPP2Zh-Comp}
\baselineskip 3.0ex
 Invariant mass $m_{Zh}$ distribution for $gg \to Zh$ production
 for $\tb=4$ and $\ma=800\gev$ at $\sqrt{s}=13\tev$.
 We plot individual contributions from the $Z$-triangle, $A$-triangle, box diagrams,
 and their interference terms.
 We use $K_{\qq}=1.31$ and
$K_{gg}=2.1$~\cite{deFlorian:2016spz, Carpenter:2016mwd}.
 }
\end{figure}

In order to see the interference effects in detail,
we split the scattering amplitude 
into the $Z$-triangle part ($\m_Z$), the $A$-triangle part ($\m_A$),
and the box part ($\m_\Box$).
The parton-level cross section becomes
\bea
\label{eq:sg:def}
\hat{\sg} &=& \frac{1}{32\pi \hat{s}} \int d \cos \theta^*
 \overline{\sum}
\big[
\left|\m_Z\right|^2
+
\left|\m_A\right|^2
+
\left|\m_\Box\right|^2
\\ \nn && ~~~~~~~~~~
+
2 \Re e \left( \m_Z \m_A^* \right)
+
2 \Re e \left( \m_Z \m_\Box^* \right)
+
2 \Re e \left( \m_A \m_\Box^* \right)
\big]
\\ \nn
&\equiv& \hat{\sg}_Z+\hat{\sg}_A+\hat{\sg}_\Box+\hat{\sg}_{ZA}+\hat{\sg}_{Z\Box}+\hat{\sg}_{A\Box},
\eea
where $\theta^*$ is the scattering angle in the center-of-mass frame
and $\overline{\sum}$ denotes the proper summation and average over helicities and colors.
Figure \ref{fig-XSPP2Zh-Comp} shows the individual contributions
 to the $m_{Zh}$ distribution of the full $gg \to Z h$ for $\tb=4$ and $\ma=800\gev$
 with $\Gm_\tot^A=2.0\tev$
at the 13 TeV LHC.
We use $K_{\qq}=1.31$ and $K_{gg}=2.1$ to match up with the updated Higgs calculation results~\cite{deFlorian:2016spz, Carpenter:2016mwd}, 
and we assume the same $K$ factor for the NP calculation.
The cross section only from the $Z$ triangle diagrams, $d \sg_Z$,
shows threshold behaviors around $m_{Zh} \simeq 2 m_t, 2 \mtp, 2 \mbp$.
The contribution from $t'$ and $b'$ is rather small
because the corresponding transition amplitude is proportional to
the axial vector coupling of the $Z$ boson, $g_A^{Zff} = - (T^f_3)_L/2$~\cite{Carpenter:2016mwd}.
For almost degenerate masses of $t'$ and $b'$,
two contributions are cancelled.
On the other hand,
$d\sg_\Box$ has large signal rate in the heavy $m_{Zh}$ range
because the opposite sign between $g_A^{Zb'b'}$ and $g_A^{Zt't'}$
is compensated by the opposite sign between $Y^h_{b'}$
and $Y^h_{t'}$.
The $d\sg_A/d m_{Zh}$ shows
a very wide resonance shape, resulting in small signal rate.
Both $d\sg_{Z\Box}$ and $d\sg_{ZA}$
yield destructive interference in the whole $m_{Zh}$ region, large enough to almost cancel $ d \sg_Z$.
The interference between the $A$-triangle and box diagrams
is constructive for $m_{Zh} \leq \ma$ while destructive for $m_{Zh} \geq \ma$,
a typical peak-dip structure~\cite{Jung:2015sna}.
In summary,
the contributions from the interference 
are as large as non-interference ones.

\begin{figure}[t]
\begin{center}
\includegraphics[width=0.7\textwidth]{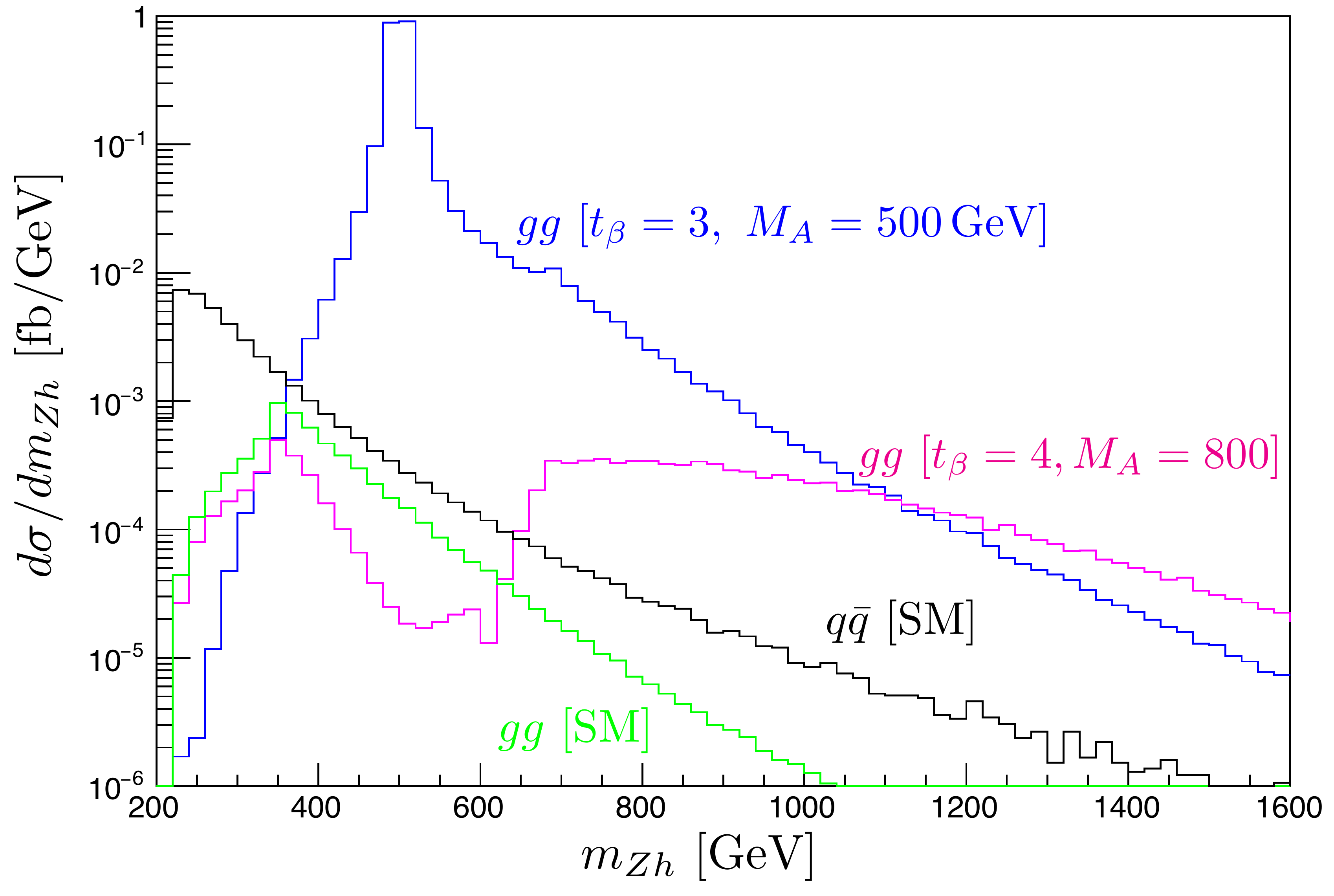}
\end{center}
\caption{\label{fig-A2Zh}
 Invariant mass $m_{Zh}$ distribution for $pp \to Zh$ process at $\sqrt{s}=13\tev$.
 We separately present $\qq$ contribution and $gg$ contributions
 in the SM and 2HDM-SM4.
 We consider two NP cases:
 (i) $\tb=3$ and $\ma=500\gev$; (ii) $\tb=4$ and $\ma=800\gev$
 in the benchmark scenario in Eq.~(\ref{eq:benchmark}).
 We use $K_{\qq}=1.31$, and
$K_{gg}=2.1$~\cite{deFlorian:2016spz, Carpenter:2016mwd}.}
\end{figure}

In Fig.~\ref{fig-A2Zh},
we present the total invariant mass distribution
in the SM and two 2HDM-SM4 cases: one with $\tb=4$ and $\ma=800\gev$ (the pink line)
and another with $\tb=3$ and $\ma=500\gev$ (the blue line).
We use $K_{\qq}=1.31$ and $K_{gg}=2.1$~\cite{deFlorian:2016spz, Carpenter:2016mwd}.
For both NP cases,
we calculate the full Feynman diagrams in Fig.~\ref{fig-Feynman-Zh}.
The total $m_{Zh}$ distribution
for $\tb=4$ and $\ma=800\gev$, of which the individual contributions are shown in Fig.~\ref{fig-XSPP2Zh-Comp},
shows a peculiar shape
with two thresholds of $\ttop$ and $\bbp$ followed by
 a very slow down hill.
This bizarre distribution is the consequence of large interference effects.
The case with $\tb=3$ and $\mhh=500\gev$  with $\Gm^A_\tot = 13.8\gev$ yields a prominent peak
over the SM main background of $\qq \to Z h$.
The LHC experiments cannot miss the peak.
Indeed the total cross section of $pp \to Z h$ for $\tb=3$ and $\ma=500\gev$ is about $24.8\pb$,
far above the upper bound on $\sigma \times \br (A \to Zh)\simeq 0.854\pb$ at $\ma=500\gev$.
Larger $\tb$ does not help to allow the model when $\ma \lsim 2 \mbp$:
(i) the resonance peak becomes more prominent because of smaller width for larger $\tb$ in this mass range
as shown in Fig.~\ref{fig-gammatot};
(ii) $\tb\gsim 4$ for $\ma<2 \mbp$ is already excluded by the $\ttau$ resonance searches as in Fig.~\ref{fig-Atautau13}.
When $\ma \gsim 2 \mbp$,
the width becomes too wide
to show a resonance peak.
We calculate the partially integrated cross section for the excess in each $m_{Zh}$ bin,
defined in Eq.(\ref{eq:Dsgnp}),
and compare with the ATLAS result of the upper bounds on $\sg \times \br(A \to Zh)$~\cite{TheATLAScollaboration:2016loc}.
We find that $\tb=4$ and $\ma = 800\gev$ is still allowed.
This conclusion is valid for larger $\tb$ and $\ma>2 \mbp$.

\subsection{Constraints from the direct searches for $H$}

\begin{figure}[h]
\begin{center}
\includegraphics[width=.6\textwidth]{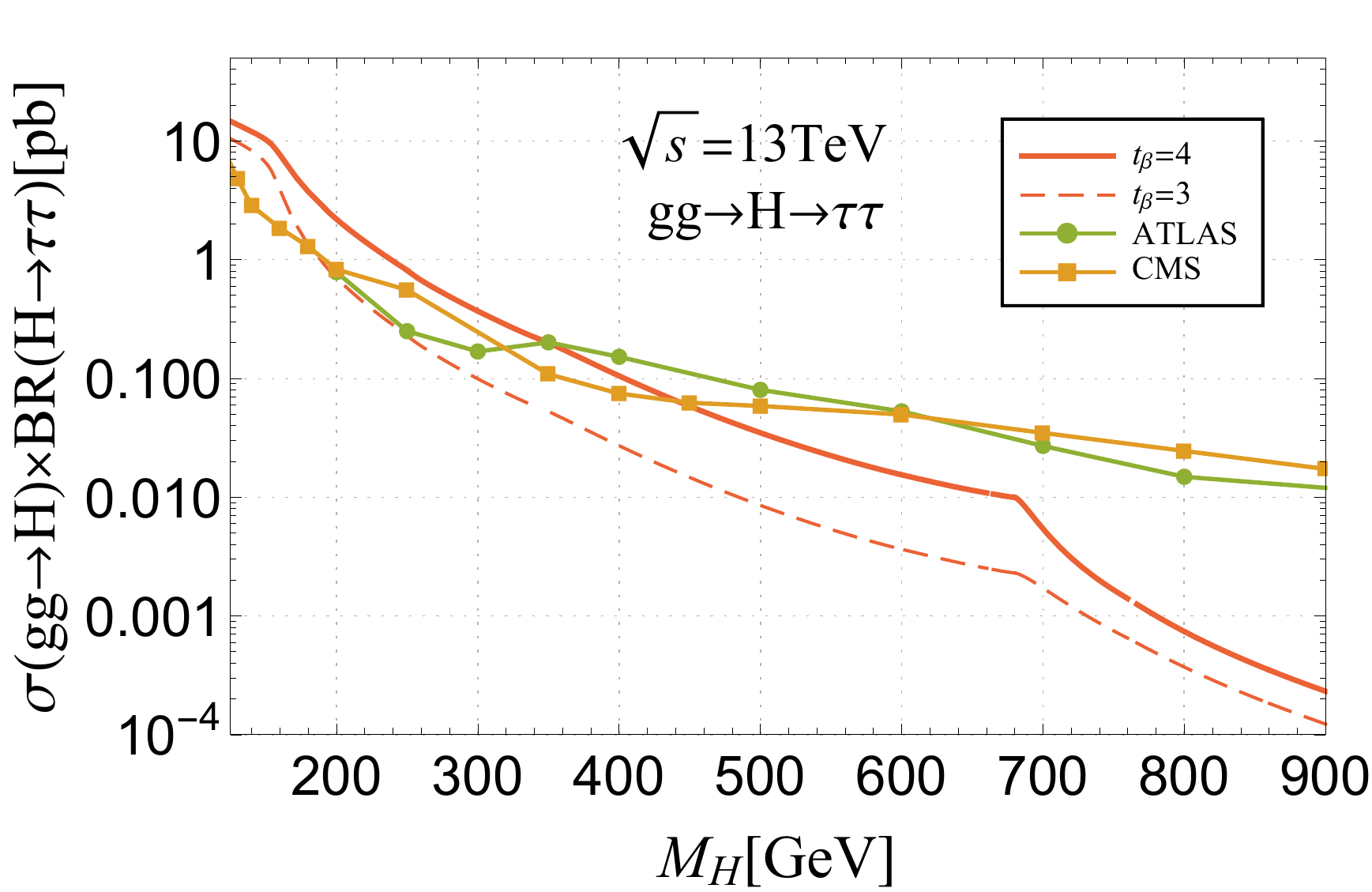}
\end{center}
\caption{\label{fig-Htautau13}
\baselineskip 3.0ex
$\sg(gg \to H)\times \br(H \to \ttau)$ 
as a function of $\mhh$ at the 13 TeV LHC.
For two cases of $\tb=3$ and $\tb=4$, 
we set $\mtp=300\gev$, $\mbp=340\gev$, $\mnup=430\gev$, and $\mtaup=380\gev$.
We apply NNLO $K$ factor to the heavy Higgs resonance production by using \texttt{HIGLU} package~\cite{Spira:1995mt}.
We also show the ATLAS and CMS 95\% C.L. upper bounds on $\sg \times \br (\phi \to \ttau)$.
}
\end{figure}

For CP-even $H$,
we first study the constraint from the resonance searches in the $\ttau$ channel
based on the ATALS and CMS
Run-2 data~\cite{Sirunyan:2018zut,Aaboud:2017sjh}.
Figure \ref{fig-Htautau13}
shows $\sg\times \br(H \to \ttau)$
in the 2HDM-SM4 at the 13 TeV LHC.
In the benchmark scenario,
we consider two cases of $\tb=3$ and $\tb=4$.
For the $K$ factor,
we take the NNLO result from \texttt{HIGLU} package~\cite{Spira:1995mt}.
Compared with the rapid drop of $\br(H \to \ttau)$ as a function of $\mhh$
in Fig.~\ref{fig-BrH},
$\sg\times \br(H \to \ttau)$ decreases slowly.
This is because of sizable cross section of the gluon fusion production of $H$
for heavy $\mhh$: see Fig.~\ref{fig-prodtot}.
The current LHC data in the $\ttau$ final states
meaningfully exclude $\mhh$,
though less than the pseudoscalar $A$.
Adopting the weaker constraint between two experiment results,
$\mhh<180\gev$ ($\mhh<350\gev$) is excluded for $\tb=3$ ($\tb=4$) at the 95\% C.L.

\begin{figure}[h]
\begin{center}
\includegraphics[width=.9\textwidth]{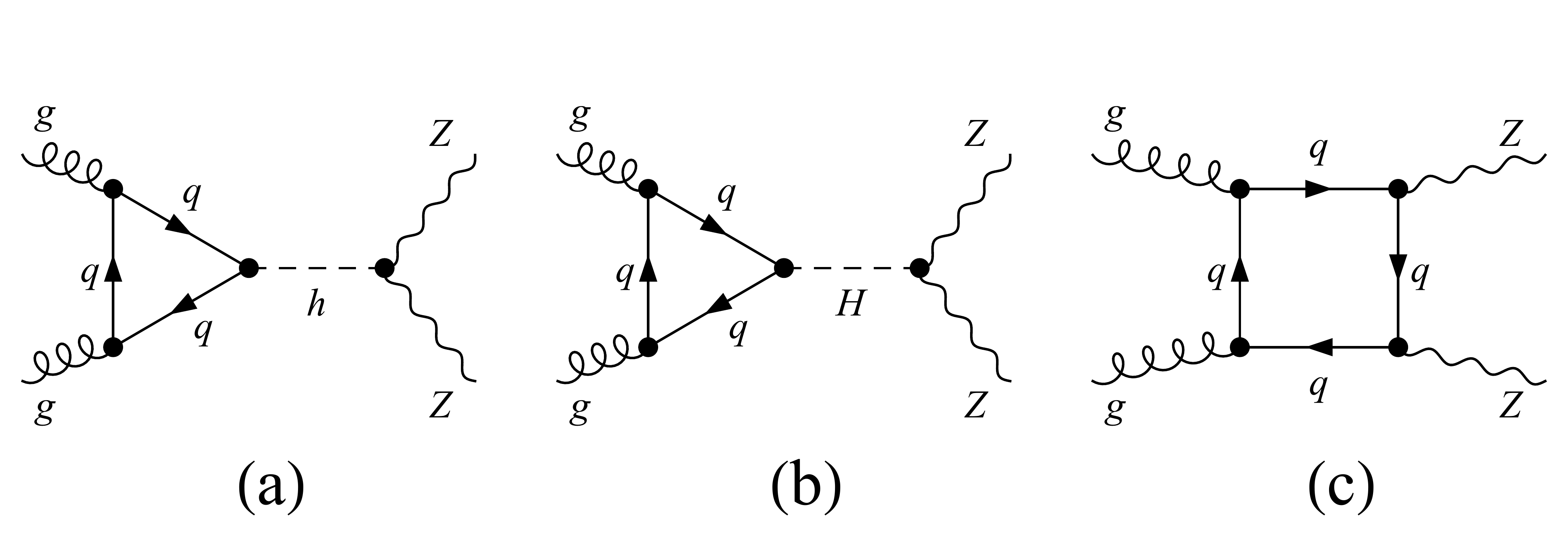}
\end{center}
\caption{\label{fig-Feynman-ZZ}
\baselineskip 3.0ex
Feynman diagrams for the $gg\to Z Z$ process in the 2HDM-SM4.
Here $q$ denotes all of the four generation quarks, including $t'$ and $b'$.
}
\end{figure}

The smoking-gun signature for the CP-even $H$ is from the $Z Z$ final state~\cite{Aaboud:2017itg,Sirunyan:2018qlb}.
In the SM, the production of a $Z$ boson pair is mainly through the $\qq$ annihilation.
The gluon fusion production via the quark loops
is sub-leading,
of which the cross section is about 10\% of the Drell-Yan process.
In the 2HDM-SM4,
the Drell-Yan process $\qq\to Z Z$ is not affected
since the CKM mixing $V_{4i}$ is extremely suppressed.
The gluon fusion process has
three kinds of Feynman diagrams as shown in Fig.~\ref{fig-Feynman-ZZ}:
(a) the triangle diagrams mediated by $h$; (b) the triangle diagrams mediated by $H$;
(c) the box diagrams.
New contributions are from
the fourth generation quarks running in the loops
and from the $H$-triangle diagram.

\begin{figure}[h]
\begin{center}
\includegraphics[width=0.7\textwidth]{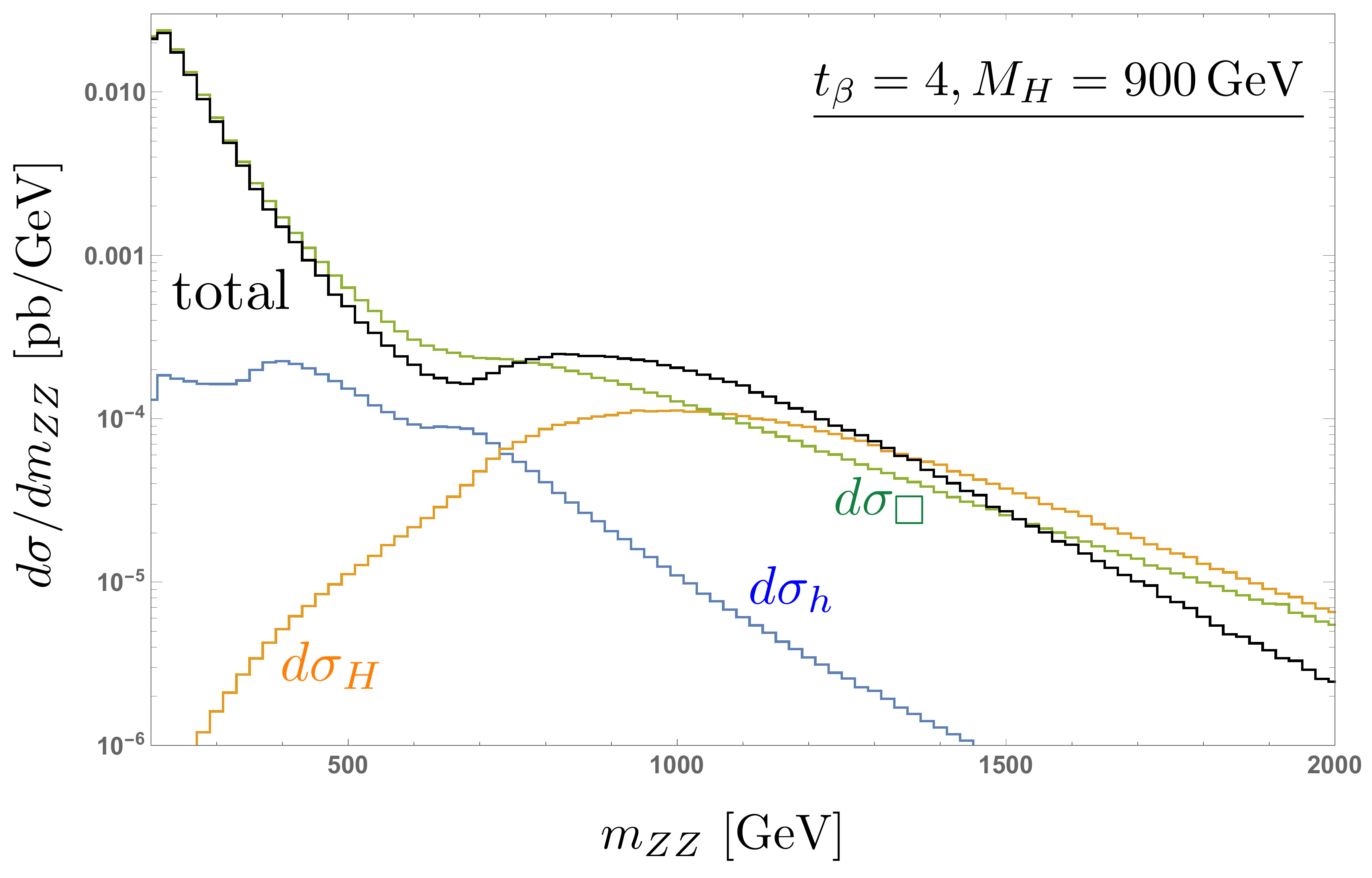}
\end{center}
\caption{\label{fig-ppZZ-diags}
The differential cross section for $gg \to Z Z$
against $m_{ZZ}$ in the 2HDM-SM4 at the 13 TeV LHC.
We set $\tb=4$ and $\mhh=900\gev$.
We show individual contributions from the $h$-triangle, $H$-triangle, and box diagrams.
We include $K_{gg}=1.8$~\cite{Caola:2015psa} and $K_{q\bar{q}}=1.5$~\cite{Cascioli:2014yka}.
}
\end{figure}

Figure \ref{fig-ppZZ-diags} shows the non-interference contributions 
to $d \sg/d m_{ZZ} (gg\to Z Z)$ from the $h$-triangle, $H$-triangle,
and box diagrams 
as a function of $m_{Z Z}$ at the 13 TeV LHC.
We set $\tb=4$ and $\mhh=900\gev$ and include $K_{gg}=1.8$ and $K_{q\bar{q}}=1.5$.
The definition of $d \sg_i$ is given in Eq.~(\ref{eq:sg:def}).
The box diagrams ($d \sg_\Box$)
yield the continuum background with monotonically decreasing slope against $m_{Z Z}$,
but the fourth generation quarks in the loop
slow the slope around the $\bbp$ threshold.
In terms of the total signal rate, the contribution from the box diagrams is dominant.
The $H$-triangle diagrams yield a very wide resonance peak around $\ma=900\gev$.
In the range of $m_{Z Z} \gsim 1\tev$,
the contribution from the $H$-triangle diagrams is as large as that from the box diagrams.
On the other hand, the $h$-triangle diagrams make a negligible contribution in the whole range of $m_{Z Z}$.
The total distribution denoted by the black line shows some deviation
from the simple sum of $d \sg_\Box + d \sg_H + d \sg_h$,
especially around $m_{Z Z} \sim 700\gev$.
Obviously the interference plays a role.

\begin{figure}[h]
\begin{center}
\includegraphics[width=0.7\textwidth]{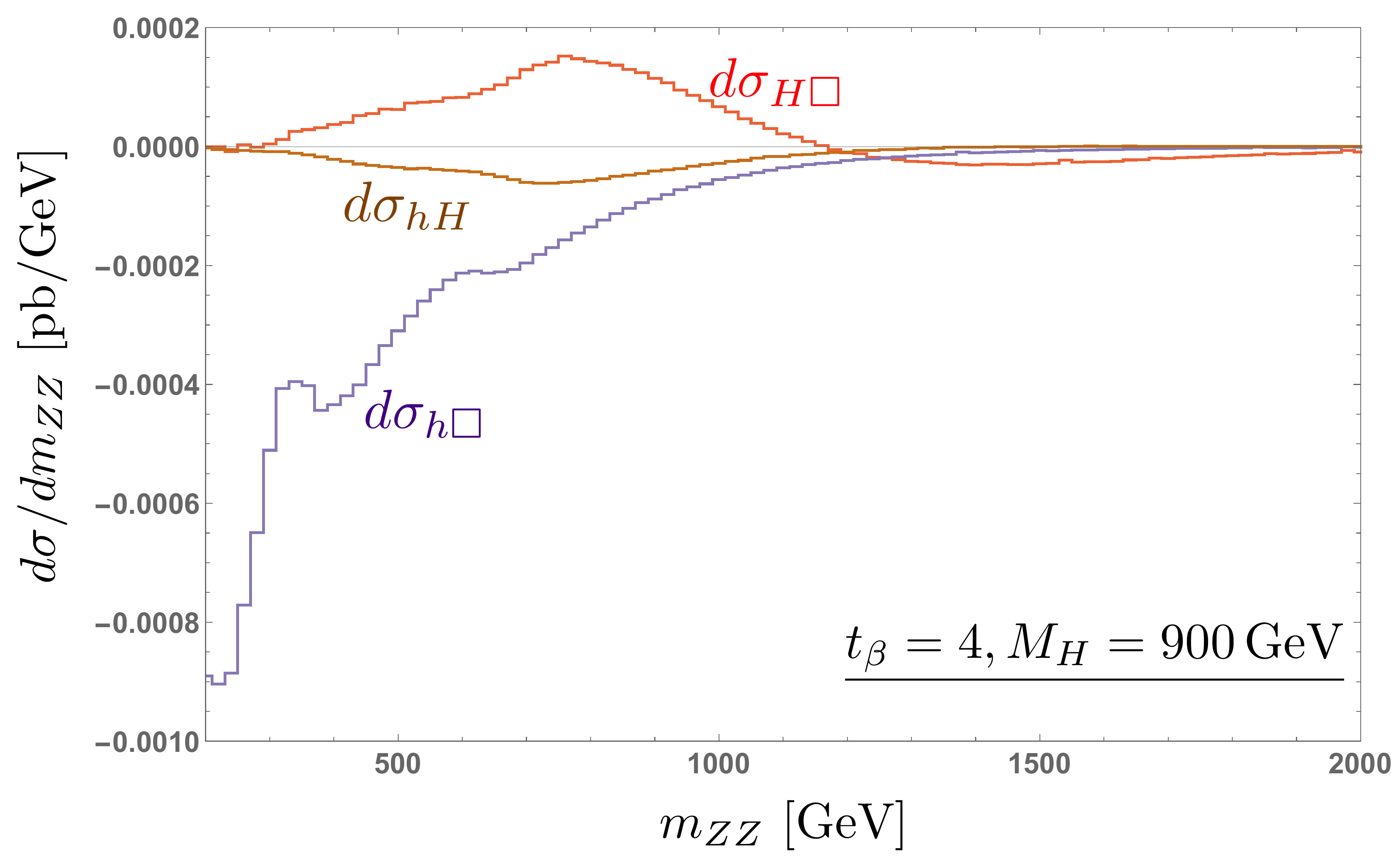}
\end{center}
\caption{\label{fig-ppZZ-diags-intf}
The differential cross section for $gg \to Z Z$
against $m_{Z Z}$ from the interference terms.
We set $\tb=4$ and $\mhh=900\gev$ at the 13 TeV LHC.
We include $K_{gg}=1.8$~\cite{Caola:2015psa} and $K_{q\bar{q}}=1.5$~\cite{Cascioli:2014yka}.
}
\end{figure}

In Fig.~\ref{fig-ppZZ-diags-intf},
we show the three kinds of interference effects, $d \sg_{ij}$ defined in Eq.~(\ref{eq:sg:def}),
that are too small to show together with non-interference $d \sg_{i}$ in a single plot.
First  the interference between the $h$-triangle and box diagrams, $\sg_{h \Box}$, is
destructive and largest.
We observe the threshold effects around $2 m_t$ and $2 \mbp$.
The interference $d \sg_{H\Box}$ shows a very asymmetric and wide peak-dip structure~\cite{Jung:2015gta},
practically like a wide peak of constructive interference.
The opposite sign between $d \sg_{h\Box}$ and $d \sg_{H\Box}$
is attributed to the opposite sign between the $h\bbp$ and $H \bbp$ couplings.
Finally $d \sg_{hH}$ is destructive but negligible.

\begin{figure}[t!]
\begin{center}
\includegraphics[width=.65\textwidth]{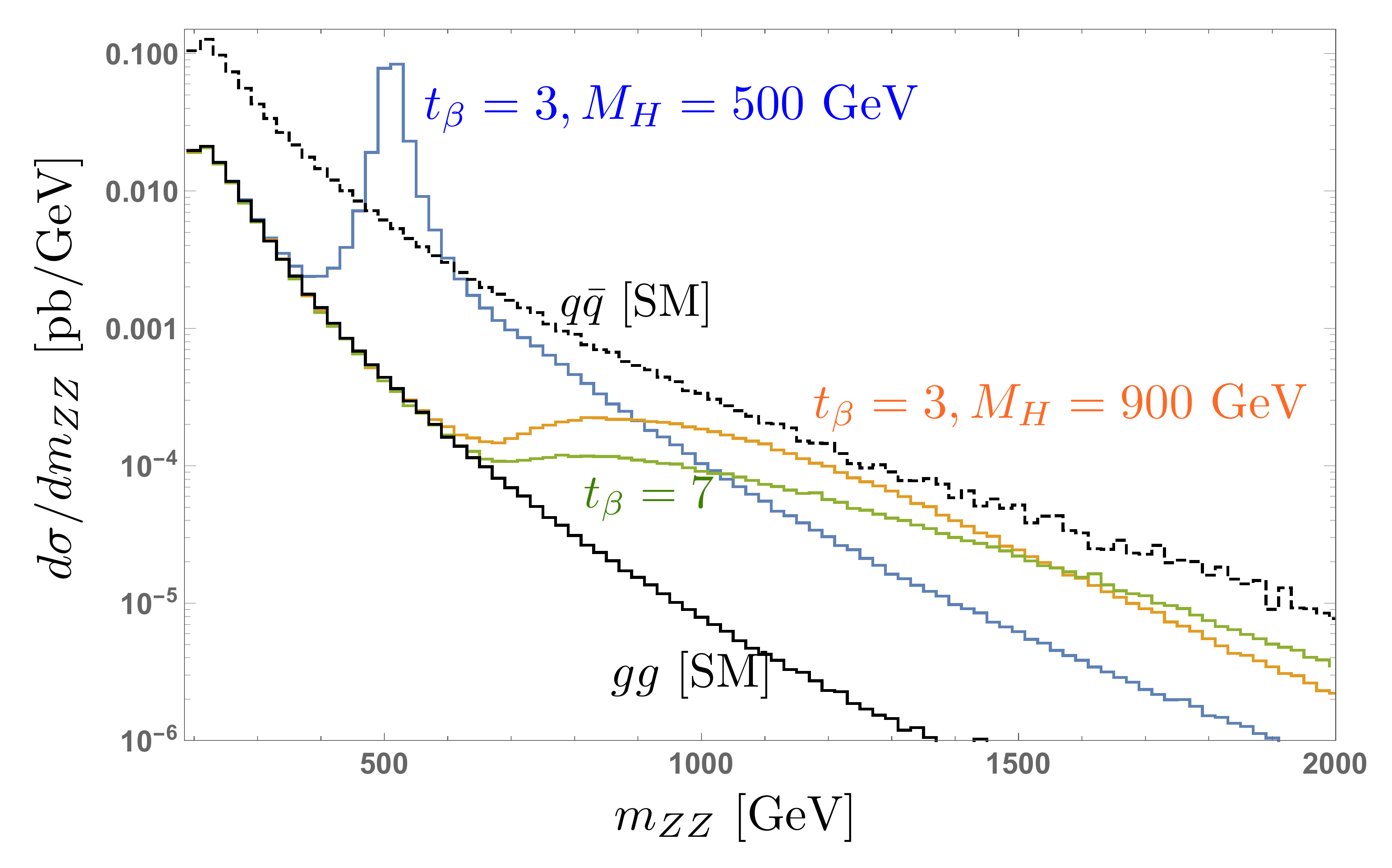}
\end{center}
\caption{\label{fig-ppZZ_13}
\baselineskip 3.0ex
The differential cross section 
against $m_{ZZ}$ in the SM and 2HDM-SM4 at the 13 TeV LHC.
We consider three NP cases: $\tb=3$ and $\mhh=500\gev$;
$\tb=4$ and $\mhh=900\gev$;
$\tb=7$ and $\mhh=900\gev$
for the benchmark point.
We include $K_{gg}=1.8$~\cite{Caola:2015psa} and $K_{q\bar{q}}=1.5$~\cite{Cascioli:2014yka}.
}
\end{figure}

Figure \ref{fig-ppZZ_13} shows the total invariant mass $m_{Z Z}$ distributions
of $\qq\to Z Z$ in the SM, $gg \to Z Z$ in the SM and the 2HDM-SM4.
We consider three NP cases:
(i) $\tb=3$ and $\mhh=500\gev$;
(ii) $\tb=4$ and $\mhh=900\gev$;
(i) $\tb=7$ and $\mhh=900\gev$.
It is clear to see that the case of $\tb=3$ and $\mhh=500\gev$,
which was allowed by the $\ttau$ constraint in Fig.~\ref{fig-Htautau13},
yields a very outstanding peak
because of relatively narrow width and large production rate of $gg \to H$.
Apparently, the current upper bound on $\sg \times \br (\phi \to Z Z)$ excludes this case.
Note that the whole parameter space with $ \mhh < 2\mbp$ 
(where 
$H$ remains as a prominent peak)
is excluded
since
larger $\tb$ yields larger production cross section of $gg\to H$ as shown in Fig.~\ref{fig-prodtot}.

When $\mhh$ is above the $\bbp$ threshold, as shown by two cases of $\tb=4,7$
with $\mhh=900\gev$,
very large width of $H$ spreads the $H$ resonance peak.
It can be seen that
the signal rate of $gg \to Z Z$ in the 2HDM-SM4
is compatible with the SM Drell-Yan production rate in the mass rage of $m_{Z Z} \gsim 1\tev$.
Comparing $\tb=4$ ($\Gm^H_\tot = 1.3\tev$) and $\tb=7$ ($\Gm^H_\tot = 3.6\tev$) cases,
we find that
the dependence on $\tb$ is not dramatically different once $\mhh$ is beyond the $\bbp$ threshold.
In order to probe this very wide resonance,
we adopt the method suggested at the end of Sec.~\ref{subsec:decay},
utilizing the 95\% C.L. upper limits on
$\sg \times \br(H \to Z Z)$ for each invariant mass bin.
Since the CMS collaboration presented the results for $\Gm_X=100\gev$,
we take the CMS results~\cite{Sirunyan:2018qlb}.

\begin{figure}[t!]
\begin{center}
\includegraphics[width=.65\textwidth]{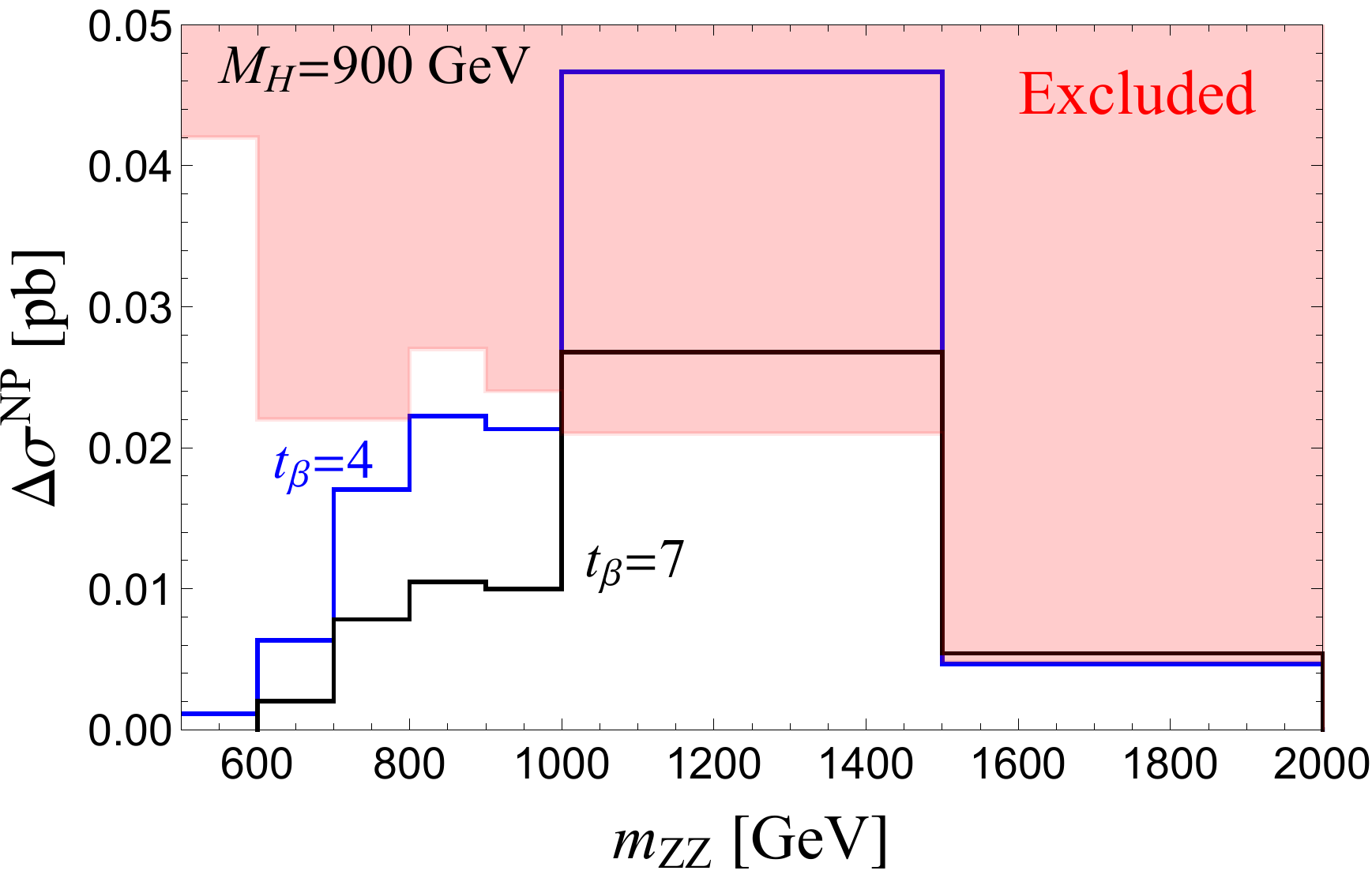}
\end{center}
\caption{\label{fig-ZZ-large-width}
Partially integrated cross section of NP effects only, $\Dt \sg^\np$
defined in Eq.~(\ref{eq:Dsgnp}),
for various $m_{Z Z}$ bins.
We set $\mhh=900\gev$ and consider $\tb=4$ and $\tb=7$.
The red region is excluded by the 95\% C.L. upper bounds on $\sg\times\br(H \to Z Z)$.}
\end{figure}

In Fig.~\ref{fig-ZZ-large-width},
we show the partially integrated cross section from NP effects only, $\Dt \sg^\np$,
for various $m_{Z Z}$ bins.
The size of bins are based on the CMS results~\cite{Sirunyan:2018qlb}.
We set $\mhh=900\gev$ and consider two cases of $\tb=4$ and $\tb=7$.
The red region is excluded by the 95\% C.L. upper bounds on $\sg\times\br(H \to Z Z)$.
Both $\tb=4$ and $\tb=7$ cases
are excluded by the bin of $m_{Z Z} \in [1,1.5]\tev$.
Larger $\tb$, apart from the breakdown of the perturbativity of $Y^H_{b'}$,
does not help to decrease the signal rate of NP
since the gluon fusion production cross section increases with larger $\tb$.
Note that we cannot increase $\mhh$ substantially further
since $\mhh=900\gev$ is almost maximally allowed value by the theoretical constraint:
see Fig.~\ref{fig-theoretical}.
In summary, the current LHC data on the $Z Z$ channel along with theoretical constraints
exclude the 2HDM-SM4 at leading order.

\section{Conclusions}
\label{sec:conclusions}

We have studied the theoretical and phenomenological constraints on the Type-II two Higgs doublet model
with a sequential fourth fermion generation in the exact wrong-sign limit, the 2HDM-SM4.
The SM fermion sector 
is extended to accommodate an additional chiral fermion generation
of which the masses are generated by the same Higgs mechanism.
Upon the absence of new fermion signals at high energy colliders,
the fourth generation fermion $F$ should be heavy, which greatly enhances
the $hFF$ couplings.
The loop induced Higgs coupling modifiers such as $\kp_g$, $\kp_\gm$, and $\kp_{Z\gm}$
become SM-like if the up-type and down-type fermions have opposite Higgs coupling modifiers:
$\kp_{\rm up}=1$ and $\kp_{\rm down}=-1$.
We call this the exact wrong-sign limit,
which can be realized in the Type-II 2HDM.

We have studied
the following constraints:
(i) the theoretical constraints on the scalar potential from
the bounded-from-below potential, unitarity, perturbativity, and the vacuum stability;
(ii) the LHC and Tevatron search bounds on the fourth generation fermion masses;
(iii) the $\bar{B} \to X_s \gm$ bound on $\mch$;
(iv) the electroweak oblique parameters $\Dt S$ and $\Dt T$;
(v) the observed Higgs coupling modifiers including the exotic Higgs decay rate;
(vi) the LEP searches for $e^+ e^- \to A h$;
(vii) the LHC scalar resonance searches in $\ttau$, $Zh$ and $Z Z$ modes.
Based on this comprehensive and thorough study,
we come to the conclusion that 
the 2HDM-SM4 is excluded at leading order.
Two features of the 2HDM-SM4 play the crucial role here,
the exact wrong-sign limit and the very large Yukawa couplings of
the down-type fourth generation fermions with $H$ and $A$, $Y^{H/A}_{b',\tau'}=\tan\beta  M_F/v$.

The exact wrong-sign limit requiring $\bt+\al =\pi/2$
constrains the model very differently from the alignment limit of
$\bt-\al =\pi/2$.
First the theoretical constraints 
do not allow the decoupling limit,
which puts the \emph{upper} bounds like
$\mhh,\ma \lsim 920\gev$ and $\mch \lsim 620\gev$.
Secondly, the constraint on 
$\tan\beta$ from the Higgs precision data 
works in the opposite way to that from the unitarity of $Y^{H/A}_{b',\tau'}=\tan\beta M_F/v$:
the observed $\kp_V (=\sba)$ at 95\% C.L.
requires $\tan\beta\gsim 3$
while the unitarity of $Y^{H/A}_{b',\tau'}$ requires $\tan\beta \lsim 8$.
The allowed value of
$\tan\beta \in [3,8]$ is not large enough to achieve
the alignment limit:
$\kp_V(=\sba)$ deviates significantly from one
and thus $\cba$ has sizable value.
Therefore, the usual 2HDM safety zone in the alignment and decoupling limit
cannot be reached in the 2HDM-SM4.

These indirect constraints
set the basic characteristics of the decays of $H$ and $A$.
Two features are to be noted.
First the sizable $\cba$ leads to significant decays of $H\to W W/Z Z/hh$ and $A \to Zh$,
which would have been absent in the alignment limit.
Above the $W W/Z Z$ ($Zh$) and below $\bbp$ threshold,
$H$ ($A$) decays dominantly into $W W/Z Z/hh$ ($Zh$).
After the $\bbp$ threshold,
the enhanced Yukawa coupling $Y^{H/A}_{b'}$ 
makes the decay of $H/A \to \bbp$ dominant.
Because of the sizable $\Gm(H/A \to gg)$
and non-negligible $\br(H/A \to \ttau)$,
the LHC direct searches in the $\ttau$ channel 
impose the strong constraint:
for $\tb=4$, $\ma < 2\mbp$ and $\mhh <350\gev$ are excluded;
for $\tb=3$, $\ma \lsim 350 \gev$ and $\mhh<180\gev$ are excluded at the 95\% C.L.

The smoking-gun signatures of $A$ and $H$
are the LHC direct searches in the $Zh$ and $Z Z$ channels, respectively.
Since their total decay widths 
are not small enough to adopt the narrow width approximation,
we performed the full calculation including the SM continuum background
without resorting to $\sg \times\br$.
The interference effects turned out to be very important.
In the $gg \to Zh$ process, they are as much as 
the non-interference ones.
We found that both $A$ and $H$
with the mass below the $\bbp$ threshold
are excluded by the current LHC data at the 95\% C.L.
We would have seen a prominent mass peak that the LHC experiment cannot miss.
Above the $\bbp$ threshold,
$\Gm^{H/A}_\tot$ becomes very large, being compatible with the mass $M_{H/A}$.
The invariant mass distributions 
spread very widely.
Beyond relying on the total event rates,
we suggested a method to utilize the upper bounds on $\sg \times \br$
on each invariant mass bin,
through the partially integrated cross section.
Although the pseudoscalar $A$ with $\ma > 2\mbp$
is allowed by the current LHC data in the $Zh$ channel,
the CP-even heavy scalar $H$
is excluded by the current LHC data in the $ZZ$ channel.

We conclude that the 2HDM-SM4 
is excluded by the combination of the theoretical and experimental constraints at leading order.
Before finishing, however,
we would like to point out that
all of the analysis are performed at leading order
and thus the conclusion may be relaxed if we consider the next-to-leading order corrections.
The sizable $\cos(\beta-\alpha)$, the multiplying factor for the $H VV$ and $A Zh$ couplings at tree level,
plays the decisive role in excluding the model.
With the fourth generation fermions running in the loop,
both $H V V$ and $AZh$ couplings 
can be meaningfully different from the tree level results.
The constraints from the LHC resonance searches in the $Z Z$ and $Zh$ final states
may be alleviated.

\acknowledgments
The work of S.K.K. was supported by NRF grants (2017K1A3A7A09016430).
The work of Z.Q. was supported by IBS under the project code, IBS-R018-D1.
The work of J.S. 
was supported by the National Research Foundation of Korea, NRF-2016R1D1A1B03932102.
Y.W.Y was supported by Basic Science Research Program through the National Research Foundation of Korea(NRF) funded by the Ministry of Education(2017R1A6A3A11036365).


\begin{thebibliography}{99}

\bibitem{Frampton:1999xi} 
  P.~H.~Frampton, P.~Q.~Hung and M.~Sher,
  Phys.\ Rept.\  {\bf 330}, 263 (2000)
  doi:10.1016/S0370-1573(99)00095-2
  [hep-ph/9903387].
  
\bibitem{Choudhury:2001hs} 
  D.~Choudhury, T.~M.~P.~Tait and C.~E.~M.~Wagner,
  Phys.\ Rev.\ D {\bf 65}, 053002 (2002)
  doi:10.1103/PhysRevD.65.053002
  [hep-ph/0109097].
  
\bibitem{Bobrowski:2009ng} 
  M.~Bobrowski, A.~Lenz, J.~Riedl and J.~Rohrwild,
  Phys.\ Rev.\ D {\bf 79}, 113006 (2009)
  doi:10.1103/PhysRevD.79.113006
  [arXiv:0902.4883 [hep-ph]].
  
\bibitem{Aad:2012tfa} 
  G.~Aad {\it et al.} [ATLAS Collaboration],
  Phys.\ Lett.\ B {\bf 716}, 1 (2012).
  
\bibitem{Chatrchyan:2012xdj} 
  S.~Chatrchyan {\it et al.} [CMS Collaboration],
  Phys.\ Lett.\ B {\bf 716}, 30 (2012).
  
\bibitem{Das:2017mnu} 
  D.~Das, A.~Kundu and I.~Saha,
  Phys.\ Rev.\ D {\bf 97}, no. 1, 011701 (2018)
  doi:10.1103/PhysRevD.97.011701
  [arXiv:1707.03000 [hep-ph]].


\bibitem{Bhattacharyya:2012tj} 
  G.~Bhattacharyya, D.~Das and P.~B.~Pal,
  Phys.\ Rev.\ D {\bf 87}, 011702 (2013)
  doi:10.1103/PhysRevD.87.011702
  [arXiv:1212.4651 [hep-ph]].

\bibitem{Carmi:2012yp} 
  D.~Carmi, A.~Falkowski, E.~Kuflik and T.~Volansky,
  JHEP {\bf 1207}, 136 (2012)
  doi:10.1007/JHEP07(2012)136
  [arXiv:1202.3144 [hep-ph]].

\bibitem{Chiang:2013ixa} 
  C.~W.~Chiang and K.~Yagyu,
  JHEP {\bf 1307}, 160 (2013)
  doi:10.1007/JHEP07(2013)160
  [arXiv:1303.0168 [hep-ph]].

\bibitem{Ferreira:2014naa} 
  P.~M.~Ferreira, J.~F.~Gunion, H.~E.~Haber and R.~Santos,
  Phys.\ Rev.\ D {\bf 89}, no. 11, 115003 (2014)
  doi:10.1103/PhysRevD.89.115003
  [arXiv:1403.4736 [hep-ph]].
       
\bibitem{Fontes:2014tga} 
  D.~Fontes, J.~C.~Romão and J.~P.~Silva,
  Phys.\ Rev.\ D {\bf 90}, no. 1, 015021 (2014)
  doi:10.1103/PhysRevD.90.015021
  [arXiv:1406.6080 [hep-ph]].

\bibitem{Hernandez-Juarez:2018uow} 
  A.~I.~Hernández-Juárez, A.~Moyotl and G.~Tavares-Velasco,
  Phys.\ Rev.\ D {\bf 98}, no. 3, 035040 (2018)
  doi:10.1103/PhysRevD.98.035040
  [arXiv:1805.00615 [hep-ph]].
  
  
\bibitem{Han:2017etg} 
  L.~Wang, R.~Shi and X.~F.~Han,
  Phys.\ Rev.\ D {\bf 96}, no. 11, 115025 (2017)
  doi:10.1103/PhysRevD.96.115025
  [arXiv:1708.06882 [hep-ph]].
    
\bibitem{Chamorro-Solano:2017toq} 
  S.~Chamorro-Solano, A.~Moyotl and M.~A.~Pérez,
  J.\ Phys.\ G {\bf 45}, no. 7, 075003 (2018)
  doi:10.1088/1361-6471/aac458
  [arXiv:1707.00100 [hep-ph]].

\bibitem{Branco:2011iw} 
  G.~C.~Branco, P.~M.~Ferreira, L.~Lavoura, M.~N.~Rebelo, M.~Sher and J.~P.~Silva,
  Phys.\ Rept.\  {\bf 516}, 1 (2012)
  doi:10.1016/j.physrep.2012.02.002
  [arXiv:1106.0034 [hep-ph]].
 
\bibitem{Adler:1969gk} 
  S.~L.~Adler,
  Phys.\ Rev.\  {\bf 177}, 2426 (1969).
  doi:10.1103/PhysRev.177.2426
  
\bibitem{Bell:1969ts} 
  J.~S.~Bell and R.~Jackiw,
  Nuovo Cim.\ A {\bf 60}, 47 (1969).
  doi:10.1007/BF02823296

\bibitem{Glashow:1976nt} 
  S.~L.~Glashow and S.~Weinberg,
  Phys.\ Rev.\ D {\bf 15}, 1958 (1977).
  doi:10.1103/PhysRevD.15.1958
  
\bibitem{Paschos:1976ay} 
  E.~A.~Paschos,
  Phys.\ Rev.\ D {\bf 15}, 1966 (1977).
  doi:10.1103/PhysRevD.15.1966
 
 
\bibitem{Chang:2015goa} 
  S.~Chang, S.~K.~Kang, J.~P.~Lee and J.~Song,
  Phys.\ Rev.\ D {\bf 92}, no. 7, 075023 (2015)
  doi:10.1103/PhysRevD.92.075023
  [arXiv:1507.03618 [hep-ph]].
  
\bibitem{Bernon:2015wef} 
  J.~Bernon, J.~F.~Gunion, H.~E.~Haber, Y.~Jiang and S.~Kraml,
  Phys.\ Rev.\ D {\bf 93}, no. 3, 035027 (2016)
  doi:10.1103/PhysRevD.93.035027
  [arXiv:1511.03682 [hep-ph]].
  
\bibitem{Pich:2009sp} 
  A.~Pich and P.~Tuzon,
  Phys.\ Rev.\ D {\bf 80}, 091702 (2009)
  doi:10.1103/PhysRevD.80.091702
  [arXiv:0908.1554 [hep-ph]].
  

\bibitem{Djouadi:2005gi} 
  A.~Djouadi,
  Phys.\ Rept.\  {\bf 457}, 1 (2008)
  doi:10.1016/j.physrep.2007.10.004
  [hep-ph/0503172].
  
\bibitem{Ferreira:2014dya} 
  P.~M.~Ferreira, R.~Guedes, M.~O.~P.~Sampaio and R.~Santos,
  JHEP {\bf 1412}, 067 (2014)
  doi:10.1007/JHEP12(2014)067
  [arXiv:1409.6723 [hep-ph]].

\bibitem{Ginzburg:2004vp} 
  I.~F.~Ginzburg and M.~Krawczyk,
  Phys.\ Rev.\ D {\bf 72}, 115013 (2005)
  doi:10.1103/PhysRevD.72.115013
  [hep-ph/0408011].

\bibitem{Dighe:2012dz} 
  A.~Dighe, D.~Ghosh, R.~M.~Godbole and A.~Prasath,
  Phys.\ Rev.\ D {\bf 85}, 114035 (2012)
  doi:10.1103/PhysRevD.85.114035
  [arXiv:1204.3550 [hep-ph]].
  
    
\bibitem{Dawson:2010jx} 
  S.~Dawson and P.~Jaiswal,
  Phys.\ Rev.\ D {\bf 82}, 073017 (2010)
  doi:10.1103/PhysRevD.82.073017
  [arXiv:1009.1099 [hep-ph]].

\bibitem{Chatrchyan:2012fp} 
  S.~Chatrchyan {\it et al.} [CMS Collaboration],
  Phys.\ Rev.\ D {\bf 86}, 112003 (2012)
  doi:10.1103/PhysRevD.86.112003
  [arXiv:1209.1062 [hep-ex]].

\bibitem{Lister:2008is} 
  A.~Lister [CDF Collaboration],
  arXiv:0810.3349 [hep-ex].
  
\bibitem{Aaltonen:2009nr} 
  T.~Aaltonen {\it et al.} [CDF Collaboration],
  Phys.\ Rev.\ Lett.\  {\bf 104}, 091801 (2010)
  doi:10.1103/PhysRevLett.104.091801
  [arXiv:0912.1057 [hep-ex]].

\bibitem{Flacco:2010rg} 
  C.~J.~Flacco, D.~Whiteson, T.~M.~P.~Tait and S.~Bar-Shalom,
  Phys.\ Rev.\ Lett.\  {\bf 105}, 111801 (2010)
  doi:10.1103/PhysRevLett.105.111801
  [arXiv:1005.1077 [hep-ph]].
  
\bibitem{Flacco:2011ym} 
  C.~J.~Flacco, D.~Whiteson and M.~Kelly,
  Phys.\ Rev.\ D {\bf 83}, 114048 (2011)
  doi:10.1103/PhysRevD.83.114048
  [arXiv:1101.4976 [hep-ph]].
  
\bibitem{Patrignani:2016xqp} 
  C.~Patrignani {\it et al.} [Particle Data Group],
  Chin.\ Phys.\ C {\bf 40}, no. 10, 100001 (2016).
  doi:10.1088/1674-1137/40/10/100001
  
\bibitem{Misiak:2017bgg} 
  M.~Misiak and M.~Steinhauser,
  Eur.\ Phys.\ J.\ C {\bf 77}, no. 3, 201 (2017).
  
\bibitem{Misiak:2017zan} 
  M.~Misiak,
  Acta Phys.\ Polon.\ B {\bf 48}, 2173 (2017).
  doi:10.5506/APhysPolB.48.2173
  
  
\bibitem{Belle:2016ufb} 
  A.~Abdesselam {\it et al.} [Belle Collaboration],
  arXiv:1608.02344 [hep-ex].
  
\bibitem{Ferreira:2014sld} 
  P.~M.~Ferreira, R.~Guedes, J.~F.~Gunion, H.~E.~Haber, M.~O.~P.~Sampaio  et al.,
  arXiv:1407.4396 [hep-ph].

\bibitem{Ivanov:2006yq} 
  I.~P.~Ivanov,
  Phys.\ Rev.\ D {\bf 75}, 035001 (2007)
  [Erratum-ibid.\ D {\bf 76}, 039902 (2007)].


\bibitem{Lee:1977eg} 
  B.~W.~Lee, C.~Quigg and H.~B.~Thacker,
  Phys.\ Rev.\ D {\bf 16}, 1519 (1977).
  doi:10.1103/PhysRevD.16.1519
  
\bibitem{Kanemura:2015ska} 
  S.~Kanemura and K.~Yagyu,
  Phys.\ Lett.\ B {\bf 751}, 289 (2015)
  doi:10.1016/j.physletb.2015.10.047
  [arXiv:1509.06060 [hep-ph]].
  
\bibitem{Arhrib:2000is} 
  A.~Arhrib,
  hep-ph/0012353.
  


\bibitem{Barroso:2013awa} 
  A.~Barroso, P.~M.~Ferreira, I.~P.~Ivanov and R.~Santos,
  JHEP {\bf 1306}, 045 (2013)
  doi:10.1007/JHEP06(2013)045
  [arXiv:1303.5098 [hep-ph]].

\bibitem{Kribs:2007nz} 
  G.~D.~Kribs, T.~Plehn, M.~Spannowsky and T.~M.~P.~Tait,
  Phys.\ Rev.\ D {\bf 76}, 075016 (2007)
  doi:10.1103/PhysRevD.76.075016
  [arXiv:0706.3718 [hep-ph]].
 
\bibitem{Grimus:2007if} 
  W.~Grimus, L.~Lavoura, O.~M.~Ogreid and P.~Osland,
  J.\ Phys.\ G {\bf 35}, 075001 (2008)
  doi:10.1088/0954-3899/35/7/075001
  [arXiv:0711.4022 [hep-ph]].

 
\bibitem{Grimus:2008nb} 
  W.~Grimus, L.~Lavoura, O.~M.~Ogreid and P.~Osland,
  Nucl.\ Phys.\ B {\bf 801}, 81 (2008)
  doi:10.1016/j.nuclphysb.2008.04.019
  [arXiv:0802.4353 [hep-ph]].
      
\bibitem{Khachatryan:2016vau} 
  G.~Aad {\it et al.} [ATLAS and CMS Collaborations],
  JHEP {\bf 1608}, 045 (2016)
  doi:10.1007/JHEP08(2016)045
  [arXiv:1606.02266 [hep-ex]].
  
  
\bibitem{Aaboud:2017vzb} 
  M.~Aaboud {\it et al.} [ATLAS Collaboration],
  JHEP {\bf 1803}, 095 (2018)
  doi:10.1007/JHEP03(2018)095
  [arXiv:1712.02304 [hep-ex]].




\bibitem{Sirunyan:2017exp} 
  A.~M.~Sirunyan {\it et al.} [CMS Collaboration],
  JHEP {\bf 1711}, 047 (2017)
  doi:10.1007/JHEP11(2017)047
  [arXiv:1706.09936 [hep-ex]].

\bibitem{Yoon:2017wul} 
  Y.~W.~Yoon, K.~Cheung, S.~K.~Kang and J.~Song,
  Phys.\ Rev.\ D {\bf 96}, no. 5, 055041 (2017)
  doi:10.1103/PhysRevD.96.055041
  [arXiv:1705.05486 [hep-ph]].

\bibitem{Aaboud:2018xdt} 
  M.~Aaboud {\it et al.} [ATLAS Collaboration],
  arXiv:1802.04146 [hep-ex].

\bibitem{CMS:2017rli} 
  CMS Collaboration [CMS Collaboration],
  CMS-PAS-HIG-16-040.

\bibitem{Chatrchyan:2013mxa} 
  S.~Chatrchyan {\it et al.} [CMS Collaboration],
  Phys.\ Rev.\ D {\bf 89}, no. 9, 092007 (2014)
  doi:10.1103/PhysRevD.89.092007
  [arXiv:1312.5353 [hep-ex]].


\bibitem{Aaboud:2017yvp} 
  M.~Aaboud {\it et al.} [ATLAS Collaboration],
  Phys.\ Rev.\ D {\bf 96}, no. 5, 052004 (2017)
  doi:10.1103/PhysRevD.96.052004
  [arXiv:1703.09127 [hep-ex]].

\bibitem{Spira:1995mt}
  M.~Spira,
  hep-ph/9510347.

\bibitem{Franceschini:2015kwy} 
  R.~Franceschini {\it et al.},
  JHEP {\bf 1603}, 144 (2016)
  doi:10.1007/JHEP03(2016)144
  [arXiv:1512.04933 [hep-ph]].

\bibitem{Djouadi:2005gj} 
  A.~Djouadi,
  Phys.\ Rept.\  {\bf 459}, 1 (2008)
  doi:10.1016/j.physrep.2007.10.005
  [hep-ph/0503173].

\bibitem{Schael:2006cr} 
  S.~Schael {\it et al.} [ALEPH and DELPHI and L3 and OPAL Collaborations and LEP Working Group for Higgs Boson Searches],
  Eur.\ Phys.\ J.\ C {\bf 47}, 547 (2006)
  doi:10.1140/epjc/s2006-02569-7
  [hep-ex/0602042].
  
\bibitem{Aad:2014vgg} 
  G.~Aad {\it et al.} [ATLAS Collaboration],
  JHEP {\bf 1411}, 056 (2014)
  doi:10.1007/JHEP11(2014)056
  [arXiv:1409.6064 [hep-ex]].

\bibitem{Khachatryan:2014wca} 
  V.~Khachatryan {\it et al.} [CMS Collaboration],
  JHEP {\bf 1410}, 160 (2014)
  doi:10.1007/JHEP10(2014)160
  [arXiv:1408.3316 [hep-ex]].

\bibitem{Aaboud:2017sjh} 
  M.~Aaboud {\it et al.} [ATLAS Collaboration],
  JHEP {\bf 1801}, 055 (2018)
  doi:10.1007/JHEP01(2018)055
  [arXiv:1709.07242 [hep-ex]].

\bibitem{Sirunyan:2018zut} 
  A.~M.~Sirunyan {\it et al.} [CMS Collaboration],
  JHEP {\bf 1809}, 007 (2018)
  doi:10.1007/JHEP09(2018)007
  [arXiv:1803.06553 [hep-ex]].


\bibitem{Sirunyan:2018qlb} 
  A.~M.~Sirunyan {\it et al.} [CMS Collaboration],
  arXiv:1804.01939 [hep-ex].
 
 
\bibitem{Aaboud:2017itg} 
  M.~Aaboud {\it et al.} [ATLAS Collaboration],
  JHEP {\bf 1803}, 009 (2018)
  doi:10.1007/JHEP03(2018)009
  [arXiv:1708.09638 [hep-ex]].
   
\bibitem{Khachatryan:2015cwa} 
  V.~Khachatryan {\it et al.} [CMS Collaboration],
  JHEP {\bf 1510}, 144 (2015)
  doi:10.1007/JHEP10(2015)144
  [arXiv:1504.00936 [hep-ex]].
  

\bibitem{Aad:2015kna} 
  G.~Aad {\it et al.} [ATLAS Collaboration],
  Eur.\ Phys.\ J.\ C {\bf 76}, no. 1, 45 (2016)
  doi:10.1140/epjc/s10052-015-3820-z
  [arXiv:1507.05930 [hep-ex]].
  
\bibitem{Khachatryan:2015lba} 
  V.~Khachatryan {\it et al.} [CMS Collaboration],
  Phys.\ Lett.\ B {\bf 748}, 221 (2015)
  doi:10.1016/j.physletb.2015.07.010
  [arXiv:1504.04710 [hep-ex]].
  
\bibitem{Aad:2015wra} 
  G.~Aad {\it et al.} [ATLAS Collaboration],
  Phys.\ Lett.\ B {\bf 744}, 163 (2015)
  doi:10.1016/j.physletb.2015.03.054
  [arXiv:1502.04478 [hep-ex]].
  
\bibitem{Miller:1999bm} 
  D.~J.~Miller, S.~Moretti, D.~P.~Roy and W.~J.~Stirling,
  Phys.\ Rev.\ D {\bf 61}, 055011 (2000)
  doi:10.1103/PhysRevD.61.055011
  [hep-ph/9906230].
  
\bibitem{Aad:2014ioa} 
  G.~Aad {\it et al.} [ATLAS Collaboration],
  Phys.\ Rev.\ Lett.\  {\bf 113}, no. 17, 171801 (2014)
  doi:10.1103/PhysRevLett.113.171801
  [arXiv:1407.6583 [hep-ex]].
  
\bibitem{CMS:2017yta} 
  CMS Collaboration [CMS Collaboration],
  CMS-PAS-HIG-17-013.
  
\bibitem{Aad:2016naf} 
  G.~Aad {\it et al.} [ATLAS Collaboration],
  Phys.\ Lett.\ B {\bf 759}, 601 (2016)
  doi:10.1016/j.physletb.2016.06.023
  [arXiv:1603.09222 [hep-ex]].

\bibitem{Aaboud:2018tqo} 
  M.~Aaboud {\it et al.} [ATLAS Collaboration],
  arXiv:1805.09299 [hep-ex].

\bibitem{deFlorian:2016spz} 
  D.~de Florian {\it et al.} [LHC Higgs Cross Section Working Group],
  doi:10.23731/CYRM-2017-002
  arXiv:1610.07922 [hep-ph].
  
\bibitem{Carpenter:2016mwd} 
  L.~M.~Carpenter, T.~Han, K.~Hendricks, Z.~Qian and N.~Zhou,
  Phys.\ Rev.\ D {\bf 95}, no. 5, 053003 (2017)
  doi:10.1103/PhysRevD.95.053003
  [arXiv:1611.05463 [hep-ph]].
  
\bibitem{TheATLAScollaboration:2016loc} 
  The ATLAS collaboration,
  ATLAS-CONF-2016-015.

\bibitem{CMS:2018xvc} 
  CMS Collaboration [CMS Collaboration],
  CMS-PAS-HIG-18-005.

\bibitem{Englert:2013vua} 
  C.~Englert, M.~McCullough and M.~Spannowsky,
  Phys.\ Rev.\ D {\bf 89}, no. 1, 013013 (2014)
  doi:10.1103/PhysRevD.89.013013
  [arXiv:1310.4828 [hep-ph]].

\bibitem{Jung:2015sna} 
  S.~Jung, Y.~W.~Yoon and J.~Song,
  Phys.\ Rev.\ D {\bf 93}, no. 5, 055035 (2016)
  doi:10.1103/PhysRevD.93.055035
  [arXiv:1510.03450 [hep-ph]].

\bibitem{Caola:2015psa} 
  F.~Caola, K.~Melnikov, R.~Röntsch and L.~Tancredi,
  Phys.\ Rev.\ D {\bf 92}, no. 9, 094028 (2015)
  doi:10.1103/PhysRevD.92.094028
  [arXiv:1509.06734 [hep-ph]].


\bibitem{Cascioli:2014yka} 
  F.~Cascioli {\it et al.},
  Phys.\ Lett.\ B {\bf 735}, 311 (2014)
  doi:10.1016/j.physletb.2014.06.056
  [arXiv:1405.2219 [hep-ph]].
  
\bibitem{Jung:2015gta} 
  S.~Jung, J.~Song and Y.~W.~Yoon,
  Phys.\ Rev.\ D {\bf 92}, no. 5, 055009 (2015)
  doi:10.1103/PhysRevD.92.055009
  [arXiv:1505.00291 [hep-ph]].
  
    
                         
\end{thebibliography}
\end{document}